\documentclass[11pt]{scrartcl}

\makeatletter
\DeclareOldFontCommand{\rm}{\normalfont\rmfamily}{\mathrm}
\DeclareOldFontCommand{\sf}{\normalfont\sffamily}{\mathsf}
\DeclareOldFontCommand{\tt}{\normalfont\ttfamily}{\mathtt}
\DeclareOldFontCommand{\bf}{\normalfont\bfseries}{\mathbf}
\DeclareOldFontCommand{\it}{\normalfont\itshape}{\mathit}
\DeclareOldFontCommand{\sl}{\normalfont\slshape}{\@nomath\sl}
\DeclareOldFontCommand{\sc}{\normalfont\scshape}{\@nomath\sc}
\makeatother

\usepackage[utf8]{inputenc}
\usepackage[english]{babel}
\usepackage{filecontents}
\usepackage{noweb}
\noweboptions{smallcode,numsubpage,nomargintag}
\usepackage{microtype}
\usepackage{xspace}
\usepackage{needspace}
\usepackage[affil-it]{authblk}

\usepackage[hidelinks]{hyperref}

\if@twoside
\else
\fi

\let\oldnwbc\nwbegincode
\def\nwbegincode{\needspace{5\baselineskip}\oldnwbc}
\def\nwendcode{\endtrivlist \endgroup}

\newcommand{\impl}[1]{\texttt{#1}}
\newcommand{\RealAlgebraic}{\texttt{Real\_algebraic}\xspace}

\title{Multithreading for the expression-dag-based number type Real\_algebraic}
\subject{Technical Report}
\author{Martin Wilhelm}
\affil{Department of Simulation and Graphics,\\ Otto von Guericke University Magdeburg}
\date{March, 2018}

\begin{document}

\maketitle
\tableofcontents


\nwenddocs{}

\section{Introduction}

Many algorithms, especially in the field of computational geometry, are based on the premise that arithmetic operations are performed exactly. Real machines, however, are based on inexact floating-point arithmetic. Various number types have been developed to close this gap by providing exact computation or, at least, ensuring exact decisions. The second approach is validated through the Exact (Geometric) Computation Paradigm, which states that making exact decisions is sufficient to guarantee the combinatorial correctness of (geometric) algorithms~\cite{yap1997}.\\

In this report we describe the implementation of an extension to the exact-decisions number type \RealAlgebraic that enables us to take advantage of multiple processing units.
\RealAlgebraic was introduced by Marc Mörig, Ivo Rössling and Stefan Schirra in 2010 with the goal of providing an easily modifiable, efficient exact-decisions number type for algebraic computations~\cite{moerig10}. In contrast to previous number types, such as \impl{leda::real}~\cite{burnikel1999} or \impl{CORE::Expr}~\cite{karamcheti1999}, it is designed as a loose framework for a variety of exchangeable components, so-called policies, making it more flexible and adaptable than its counterparts.\\

\RealAlgebraic uses a lazy evaluation procedure. When the user invokes an arithmetic operation, it first tries to exactly compute the result locally. If this is not possible, it builds an expression dag. When an expression dag is built, the expression does not get evaluated until its value is requested through a comparison\footnote{The evaluation can also be started manually to guarantee a certain error bound, but this is not the use case the number type is designed for.}. Then a floating-point filter is invoked and only if the error bound of the filter is not sufficient, the expression gets recomputed with bigfloats up until a sufficiently small error is reached to make an exact decision, i.e., to separate it from zero or to decide that the result is zero through a separation bound (cf.~\cite{fortune1993,schirra2009}).

A very fundamental policy is the \impl{ExpressionDagPolicy}, which regulates the design and the evaluation process of the underlying expression dag. The main part of this policy is the design of the underlying node class. We describe a new kind of dag node based on the standard \impl{sdag\_node}. We include the complete code for our new dag node in this report. However, we will not include any code regarding other policies or \RealAlgebraic itself. For more detailed information on the design and the source code of \RealAlgebraic, see~\cite{moerig10,moerig15}.\\

Our new node type with multithreaded evaluation will be called \impl{mtdag\_node}. To parallelize the evaluation we are going to use the C++11 multithreading environment, since it is platform independent, does not require any additional libraries and fits nicely into the programming logic~\cite{isocpp11}. However, we will as well use the boost thread library to gain some efficiency when implementing a thread manager~\cite{boostlockfree}.

In Section~\ref{sss:threadmanagement}, we will create a class managing the number of concurrent threads. In Section~\ref{sss:threadapplication} we describe the new node class.

\section{Thread Management}
\label{sss:threadmanagement}

In this section we describe the design and the implementation of a thread manager for the new dag node.\\

While in theory arbitrarily many processors can be used, a real machine has only a small number of processing units. If the number of threads exceeds the number of processing units, concurrency will be simulated by constantly switching threads. If the number of context switches during a computation is high, this can lead to serious performance issues. Therefore it is important to keep the number of threads created at any point of the computation low.\\

Since we want to abstract from limitations in the number of threads created, we create a thread pool, called \impl{Thread\_manager}, which can receive an arbitrary number of tasks that may potentially be executed in parallel.

\paragraph{Outline}
\ \\
The \impl{Thread\_manager} class follows an on-demand approach in the sense that, instead of maintaining a constant sized thread pool, we will create more threads as soon as there are enough unfinished tasks for this to pay off. However, we only create new threads up to a certain threshold to prevent excessive context switching.

This approach has proven to be superior to a permanent thread pool, since the structure of many expression dags is not suited for an efficient use of multiple threads. Even when we have a beneficial structure, for example a perfectly balanced tree, we get faster through eliminating unused threads at the end of the computation, when we have few, but expensive operations near the root of the tree.\\

From an implementation point of view, we put some effort in making the thread manager lock-free, i.e., we avoid using mutexes but instead exploit atomic operations when operating on shared data. In particular we partly give up control over the generated threads, which leads to the following condition for usage of \impl{Thread\_manager}:
\begin{itemize}
  \item The \emph{caller}, which is the one providing tasks to the thread manager, must assure that no tasks are open when the destructor of the thread manager is called.
\end{itemize}

This is a consequence of the fact that we neither have nor want to implement waiting mechanisms inside the class. In addition, it means that the class does not provide any mechanisms to notify the caller, when its tasks are finished.

\nwfilename{techreport.nw}\nwbegincode{1}\sublabel{NWwxjkf-47sd8i-1}\nwmargintag{{\nwtagstyle{}\subpageref{NWwxjkf-47sd8i-1}}}\moddef{Thread\_manager.hpp~{\nwtagstyle{}\subpageref{NWwxjkf-47sd8i-1}}}\endmoddef\nwstartdeflinemarkup\nwenddeflinemarkup
#ifndef RA_THREAD_MANAGER_HPP
#define RA_THREAD_MANAGER_HPP

#if (defined RA_CXX11 && defined RA_WITH_BOOST)

\LA{}tm includes~{\nwtagstyle{}\subpageref{NWwxjkf-2sbsH1-1}}\RA{}
\LA{}tm debug includes~{\nwtagstyle{}\subpageref{NWwxjkf-3yauSd-1}}\RA{}

class Thread_manager\{ 

private:

\LA{}tm members~{\nwtagstyle{}\subpageref{NWwxjkf-23cSgs-1}}\RA{}
\LA{}tm thread and task handling~{\nwtagstyle{}\subpageref{NWwxjkf-lCGr5-1}}\RA{}
\LA{}tm constructors and destructors~{\nwtagstyle{}\subpageref{NWwxjkf-2SHrAR-1}}\RA{}
\LA{}tm guard~{\nwtagstyle{}\subpageref{NWwxjkf-2q8kM5-1}}\RA{}

public:

\LA{}tm add task~{\nwtagstyle{}\subpageref{NWwxjkf-42cCML-1}}\RA{}
\LA{}tm instantiation~{\nwtagstyle{}\subpageref{NWwxjkf-4gZXMR-1}}\RA{}

\};

#endif//RA_CXX11 && RA_WITH_BOOST

#endif//RA_THREAD_MANAGER_HPP
\nwnotused{Thread\_manager.hpp}\nwendcode{}\nwbegindocs{2}%

For storing the tasks we use a \impl{boost::lockfree::queue}~\cite{boostlockfree}, which is why having boost is a hard requirement for using the \impl{Thread\_manager} and, consequently, the multithreaded dag node.

\paragraph{Singleton creation}
\ \\
The \impl{Thread\_manager} is implemented as a singleton, i.e., there can always be at most one instance of the class which then will be retained until the execution stops. This prevents the class object from getting created again each time a computation is executed.

To achieve this we create our instance in a private static member variable
\nwenddocs{}\nwbegincode{3}\sublabel{NWwxjkf-23cSgs-1}\nwmargintag{{\nwtagstyle{}\subpageref{NWwxjkf-23cSgs-1}}}\moddef{tm members~{\nwtagstyle{}\subpageref{NWwxjkf-23cSgs-1}}}\endmoddef\nwstartdeflinemarkup\nwusesondefline{\\{NWwxjkf-47sd8i-1}}\nwprevnextdefs{\relax}{NWwxjkf-23cSgs-2}\nwenddeflinemarkup
static Thread_manager *instance;
\nwalsodefined{\\{NWwxjkf-23cSgs-2}\\{NWwxjkf-23cSgs-3}\\{NWwxjkf-23cSgs-4}}\nwused{\\{NWwxjkf-47sd8i-1}}\nwendcode{}\nwbegindocs{4}%

which we then make accessible from outside the class through a public getter function. In the getter function we create an instance of our class, when it is requested for the first time.
\nwenddocs{}\nwbegincode{5}\sublabel{NWwxjkf-4gZXMR-1}\nwmargintag{{\nwtagstyle{}\subpageref{NWwxjkf-4gZXMR-1}}}\moddef{tm instantiation~{\nwtagstyle{}\subpageref{NWwxjkf-4gZXMR-1}}}\endmoddef\nwstartdeflinemarkup\nwusesondefline{\\{NWwxjkf-47sd8i-1}}\nwenddeflinemarkup
static Thread_manager* get_instance() \{
  \LA{}tm guard instantiation~{\nwtagstyle{}\subpageref{NWwxjkf-3L7KKX-1}}\RA{}
  
  if (instance == nullptr) instance = new Thread_manager();
  return instance;
\}
\nwused{\\{NWwxjkf-47sd8i-1}}\nwendcode{}\nwbegindocs{6}%

We need to make sure that the instance gets deleted when the execution stops. For this we use a (private) guard class, which will delete the instance as soon as its destructor gets called.
\nwenddocs{}\nwbegincode{7}\sublabel{NWwxjkf-2q8kM5-1}\nwmargintag{{\nwtagstyle{}\subpageref{NWwxjkf-2q8kM5-1}}}\moddef{tm guard~{\nwtagstyle{}\subpageref{NWwxjkf-2q8kM5-1}}}\endmoddef\nwstartdeflinemarkup\nwusesondefline{\\{NWwxjkf-47sd8i-1}}\nwenddeflinemarkup
class Thread_manager_guard \{
public:
  ~Thread_manager_guard() \{
    if (instance != nullptr) \{
      delete instance;
      instance = nullptr;
    \}
  \}
\};
\nwused{\\{NWwxjkf-47sd8i-1}}\nwendcode{}\nwbegindocs{8}%

We then create an instance of this guard class as soon as we get the first call to \impl{get\_instance}.

\nwenddocs{}\nwbegincode{9}\sublabel{NWwxjkf-3L7KKX-1}\nwmargintag{{\nwtagstyle{}\subpageref{NWwxjkf-3L7KKX-1}}}\moddef{tm guard instantiation~{\nwtagstyle{}\subpageref{NWwxjkf-3L7KKX-1}}}\endmoddef\nwstartdeflinemarkup\nwusesondefline{\\{NWwxjkf-4gZXMR-1}}\nwenddeflinemarkup
static Thread_manager_guard guard;
\nwused{\\{NWwxjkf-4gZXMR-1}}\nwendcode{}\nwbegindocs{10}%

By this we have guaranteed that
\begin{itemize}
  \item There is always at most one instance of \impl{Thread\_manager}
  \item The instance does not get destroyed after each computation.
  \item The instance gets destroyed when the program execution stops.
\end{itemize}

\paragraph{Task and thread management}
\ \\
We are going to store incoming tasks in a \impl{boost::lockfree::queue}.
\nwenddocs{}\nwbegincode{11}\sublabel{NWwxjkf-23cSgs-2}\nwmargintag{{\nwtagstyle{}\subpageref{NWwxjkf-23cSgs-2}}}\moddef{tm members~{\nwtagstyle{}\subpageref{NWwxjkf-23cSgs-1}}}\plusendmoddef\nwstartdeflinemarkup\nwusesondefline{\\{NWwxjkf-47sd8i-1}}\nwprevnextdefs{NWwxjkf-23cSgs-1}{NWwxjkf-23cSgs-3}\nwenddeflinemarkup
boost::lockfree::queue<std::function<void()>*> tasks;
\nwused{\\{NWwxjkf-47sd8i-1}}\nwendcode{}\nwbegindocs{12}%

This is a queue that is both multithread-safe and lock-free, which means it can handle concurrent pop/pop, pop/push and push/push situations without compromising its internal structure \emph{and} without blocking any reader or writer thread in its execution.

The tasks are given as a function object without arguments. So the caller has to bind potential arguments to the function before assigning a task to the \impl{Thread\_manager}. We need to include the respective headers for the queue and the functional.

\nwenddocs{}\nwbegincode{13}\sublabel{NWwxjkf-2sbsH1-1}\nwmargintag{{\nwtagstyle{}\subpageref{NWwxjkf-2sbsH1-1}}}\moddef{tm includes~{\nwtagstyle{}\subpageref{NWwxjkf-2sbsH1-1}}}\endmoddef\nwstartdeflinemarkup\nwusesondefline{\\{NWwxjkf-47sd8i-1}}\nwprevnextdefs{\relax}{NWwxjkf-2sbsH1-2}\nwenddeflinemarkup
#include <functional>
#include <boost/lockfree/queue.hpp>
\nwalsodefined{\\{NWwxjkf-2sbsH1-2}\\{NWwxjkf-2sbsH1-3}}\nwused{\\{NWwxjkf-47sd8i-1}}\nwendcode{}\nwbegindocs{14}%

Now launched threads will be pulling tasks from the queue until it is empty. When this is the case they are going to automatically destroy themselves. So each new thread will be started with the \impl{execute\_tasks} method.
\nwenddocs{}\nwbegincode{15}\sublabel{NWwxjkf-lCGr5-1}\nwmargintag{{\nwtagstyle{}\subpageref{NWwxjkf-lCGr5-1}}}\moddef{tm thread and task handling~{\nwtagstyle{}\subpageref{NWwxjkf-lCGr5-1}}}\endmoddef\nwstartdeflinemarkup\nwusesondefline{\\{NWwxjkf-47sd8i-1}}\nwprevnextdefs{\relax}{NWwxjkf-lCGr5-2}\nwenddeflinemarkup
void execute_tasks() \{
  std::function<void()>* currtask;
  while(tasks.pop(currtask))
  \{
    \LA{}tm execute task~{\nwtagstyle{}\subpageref{NWwxjkf-fJsF4-1}}\RA{}
  \}
  \LA{}tm destroy thread~{\nwtagstyle{}\subpageref{NWwxjkf-AjUMk-1}}\RA{}
\}
\nwalsodefined{\\{NWwxjkf-lCGr5-2}\\{NWwxjkf-lCGr5-3}}\nwused{\\{NWwxjkf-47sd8i-1}}\nwendcode{}\nwbegindocs{16}%

The \impl{pop} method of the queue will return false, if it was not able to pop an element, hence the queue was empty at the time the call started. To keep track of the current number of threads, as well as the current number of tasks we store two atomic counters in addition to the queue itself.

\nwenddocs{}\nwbegincode{17}\sublabel{NWwxjkf-23cSgs-3}\nwmargintag{{\nwtagstyle{}\subpageref{NWwxjkf-23cSgs-3}}}\moddef{tm members~{\nwtagstyle{}\subpageref{NWwxjkf-23cSgs-1}}}\plusendmoddef\nwstartdeflinemarkup\nwusesondefline{\\{NWwxjkf-47sd8i-1}}\nwprevnextdefs{NWwxjkf-23cSgs-2}{NWwxjkf-23cSgs-4}\nwenddeflinemarkup
std::atomic<unsigned int> thread_count;
std::atomic<unsigned int> task_count;
\nwused{\\{NWwxjkf-47sd8i-1}}\nwendcode{}\nwbegindocs{18}%

Note that the queue does not provide any method to check its size, since there is no way of \emph{knowing} the exact size during concurrent pop/push operations. Atomic counters ensure that each increment or decrement operation is executed atomically, which means that no other thread can modify the data while the operation takes place. For native integer data types this should also be done lock-free. We need to include the respective \impl{atomic} header\footnote{We do not really need to include the header, since it should be included by the lockfree queue as well, but this adds to code stability.}.
\nwenddocs{}\nwbegincode{19}\sublabel{NWwxjkf-2sbsH1-2}\nwmargintag{{\nwtagstyle{}\subpageref{NWwxjkf-2sbsH1-2}}}\moddef{tm includes~{\nwtagstyle{}\subpageref{NWwxjkf-2sbsH1-1}}}\plusendmoddef\nwstartdeflinemarkup\nwusesondefline{\\{NWwxjkf-47sd8i-1}}\nwprevnextdefs{NWwxjkf-2sbsH1-1}{NWwxjkf-2sbsH1-3}\nwenddeflinemarkup
#include <atomic>
\nwused{\\{NWwxjkf-47sd8i-1}}\nwendcode{}\nwbegindocs{20}%

Now when we execute a task in our \impl{execute\_tasks} method, we are going to decrease the task counter. Afterwards we call the associated function and delete the function object.

\nwenddocs{}\nwbegincode{21}\sublabel{NWwxjkf-fJsF4-1}\nwmargintag{{\nwtagstyle{}\subpageref{NWwxjkf-fJsF4-1}}}\moddef{tm execute task~{\nwtagstyle{}\subpageref{NWwxjkf-fJsF4-1}}}\endmoddef\nwstartdeflinemarkup\nwusesondefline{\\{NWwxjkf-lCGr5-1}}\nwenddeflinemarkup
--task_count;
currtask->operator()();
delete currtask;
\nwused{\\{NWwxjkf-lCGr5-1}}\nwendcode{}\nwbegindocs{22}%

When we exit the loop and are about to destroy our current thread, we need to decrement the thread counter. Since this involves some safety measures, it is done in a separate method, we will describe later on.
\nwenddocs{}\nwbegincode{23}\sublabel{NWwxjkf-AjUMk-1}\nwmargintag{{\nwtagstyle{}\subpageref{NWwxjkf-AjUMk-1}}}\moddef{tm destroy thread~{\nwtagstyle{}\subpageref{NWwxjkf-AjUMk-1}}}\endmoddef\nwstartdeflinemarkup\nwusesondefline{\\{NWwxjkf-lCGr5-1}}\nwenddeflinemarkup
dec_thread_count();
\nwused{\\{NWwxjkf-lCGr5-1}}\nwendcode{}\nwbegindocs{24}%

As mentioned in the outline, we will create new threads on demand, up to a certain maximum. For this we define two constants. The constant \impl{MAX\_THREADS} determines the number of concurrent threads at which the creation of new threads will stop. Consequently, it can be seen as the maximum number of threads running at any time during the execution\footnote{This is not entirely correct, as described later on}. The prefered value of this constant depends heavily on the number of available processing units on the target machine and should be adjusted accordingly. For maximal efficiency, this constant should be about the same size as the number of available processing units.

The second constant, \impl{TASK\_THRESHOLD}, determines the number of queued, unfinished tasks, at which a new thread will be created. If this number is too low, this may result in continous creation and destruction of new threads and therefore negatively influence the performance. On the other hand, in our previous tests, we could not see a notable difference in performance when increasing the threshold above five up to one hundred.
\nwenddocs{}\nwbegincode{25}\sublabel{NWwxjkf-23cSgs-4}\nwmargintag{{\nwtagstyle{}\subpageref{NWwxjkf-23cSgs-4}}}\moddef{tm members~{\nwtagstyle{}\subpageref{NWwxjkf-23cSgs-1}}}\plusendmoddef\nwstartdeflinemarkup\nwusesondefline{\\{NWwxjkf-47sd8i-1}}\nwprevnextdefs{NWwxjkf-23cSgs-3}{\relax}\nwenddeflinemarkup
static constexpr int MAX_THREADS = 4;
static constexpr int TASK_THRESHOLD = 5;
\nwused{\\{NWwxjkf-47sd8i-1}}\nwendcode{}\nwbegindocs{26}%
As we will see soon, due to the non-deterministic execution order, these constants do \emph{neither} imply that there will never be more threads active than \impl{MAX\_THREADS}, \emph{nor} that a number of tasks beyond \impl{TASK\_THRESHOLD} always leads to the creation of a new thread, \emph{nor} that the \impl{TASK\_THRESHOLD} must be reached for a new thread to be created. So these constants should always be seen as soft conditions for task and thread management.\\

Now that we defined our conditions for thread creation, we describe the methods, in which new threads will be created. This is primarily done, whenever a new task will be added to the thread manager. We define our public interface for task creation as follows
\nwenddocs{}\nwbegincode{27}\sublabel{NWwxjkf-42cCML-1}\nwmargintag{{\nwtagstyle{}\subpageref{NWwxjkf-42cCML-1}}}\moddef{tm add task~{\nwtagstyle{}\subpageref{NWwxjkf-42cCML-1}}}\endmoddef\nwstartdeflinemarkup\nwusesondefline{\\{NWwxjkf-47sd8i-1}}\nwenddeflinemarkup
void add_task(std::function<void()>* pt) \{
  ++task_count;
  tasks.push(pt);

  \LA{}tm conditioned launch new thread~{\nwtagstyle{}\subpageref{NWwxjkf-1cLBPM-1}}\RA{}
\}
\nwused{\\{NWwxjkf-47sd8i-1}}\nwendcode{}\nwbegindocs{28}%

We increase the (atomic) task counter and add the new task to our queue. Note that we first increase the counter and only afterwards store the task in the queue, while in the \impl{execute\_tasks} method, we first remove an element from the queue and afterwards decrease the task counter. Therefore the value of the task counter is always greater or equal to the number of tasks stored inside the queue. In particular, \impl{task\_count} is zero \emph{only if} the number of tasks inside the queue is zero.\\

After adding the new task to the queue, we check, whether we should create a new thread
\nwenddocs{}\nwbegincode{29}\sublabel{NWwxjkf-1cLBPM-1}\nwmargintag{{\nwtagstyle{}\subpageref{NWwxjkf-1cLBPM-1}}}\moddef{tm conditioned launch new thread~{\nwtagstyle{}\subpageref{NWwxjkf-1cLBPM-1}}}\endmoddef\nwstartdeflinemarkup\nwusesondefline{\\{NWwxjkf-42cCML-1}}\nwenddeflinemarkup
if (thread_count == 0 || task_count > TASK_THRESHOLD) \{
  \LA{}tm launch thread if not max~{\nwtagstyle{}\subpageref{NWwxjkf-3BVqFP-1}}\RA{}
\}
\nwused{\\{NWwxjkf-42cCML-1}}\nwendcode{}\nwbegindocs{30}%
Obviously we need to have at least one thread running, whenever there are unfinished tasks inside our queue. Therefore we want to create a new thread, when \impl{thread\_count} is zero. Secondly we want to create a new thread, when the number of tasks waiting in our queue gets high and therefore \impl{task\_count} surpasses the \impl{TASK\_THRESHOLD}. Since the number of tasks inside the queue can be smaller than the value of \impl{task\_count}, we cannot guarantee that the threshold is actually reached, when a new thread is created.\\

Then we create a new thread unless the maximum number of threads is reached.
\nwenddocs{}\nwbegincode{31}\sublabel{NWwxjkf-3BVqFP-1}\nwmargintag{{\nwtagstyle{}\subpageref{NWwxjkf-3BVqFP-1}}}\moddef{tm launch thread if not max~{\nwtagstyle{}\subpageref{NWwxjkf-3BVqFP-1}}}\endmoddef\nwstartdeflinemarkup\nwusesondefline{\\{NWwxjkf-1cLBPM-1}}\nwenddeflinemarkup
if (thread_count < MAX_THREADS) \{
  launch_thread();
\}
\nwused{\\{NWwxjkf-1cLBPM-1}}\nwendcode{}\nwbegindocs{32}%
Note that this does not completely prevent the number of threads to surpass \impl{MAX\_THREADS}. It is possible that several calls to \impl{add\_task} reach this line simultaneously and therefore pass the check before increasing the thread counter.\\

When we want to create a new thread, we call the \impl{launch\_thread} method.
\nwenddocs{}\nwbegincode{33}\sublabel{NWwxjkf-lCGr5-2}\nwmargintag{{\nwtagstyle{}\subpageref{NWwxjkf-lCGr5-2}}}\moddef{tm thread and task handling~{\nwtagstyle{}\subpageref{NWwxjkf-lCGr5-1}}}\plusendmoddef\nwstartdeflinemarkup\nwusesondefline{\\{NWwxjkf-47sd8i-1}}\nwprevnextdefs{NWwxjkf-lCGr5-1}{NWwxjkf-lCGr5-3}\nwenddeflinemarkup
inline void launch_thread() \{
  ++thread_count;
  std::thread t(&Thread_manager::execute_tasks,this);
  t.detach();
\}
\nwused{\\{NWwxjkf-47sd8i-1}}\nwendcode{}\nwbegindocs{34}%
In \impl{launch\_thread} the value of the thread counter will be increased and a new thread will be created. The newly created thread will then be detached, which means it will run independently from the \impl{Thread\_manager} object until the executed method, \impl{execute\_tasks} comes to an end. So after leaving this method, we completely give up control over this thread. Therefore it is important (cf.\ the outline of this section) that the caller ensures that all tasks have finished before the \impl{Thread\_manager} object gets destroyed\footnote{There is a very unlikely, but (maybe?) possible scenario, in which all tasks are finished, the \impl{Thread\_manager} gets destroyed and then one dying thread tries to pop a task from the (already destroyed) queue. Whether this situation can happen, has any implications, and how to prevent this from happening still needs further investigation.}.\\

We increase the thread counter before a new thread gets created, whereas in \impl{execute\_tasks} we exit the loop (and therefore effectively kill the thread) before we decrease the thread counter by calling \impl{dec\_thread\_count}. Therefore, as with \impl{task\_count}, the value of \impl{thread\_count} is always greater or equal to the number of threads that are (effectively) alive. So whenever \impl{thread\_count} is zero, the number of threads alive is zero as well.\\

However, we cannot reverse the implication. That means, checking whether \impl{thread\_count} is greater than zero, as done in \impl{add\_task}, is \emph{not} sufficient to ensure that there are any threads alive. Therefore we must check this condition again, after decreasing the thread counter.
\nwenddocs{}\nwbegincode{35}\sublabel{NWwxjkf-lCGr5-3}\nwmargintag{{\nwtagstyle{}\subpageref{NWwxjkf-lCGr5-3}}}\moddef{tm thread and task handling~{\nwtagstyle{}\subpageref{NWwxjkf-lCGr5-1}}}\plusendmoddef\nwstartdeflinemarkup\nwusesondefline{\\{NWwxjkf-47sd8i-1}}\nwprevnextdefs{NWwxjkf-lCGr5-2}{\relax}\nwenddeflinemarkup
inline void dec_thread_count() \{
  if (--thread_count == 0 && task_count > 0) \{
    launch_thread();
  \}
\}
\nwused{\\{NWwxjkf-47sd8i-1}}\nwendcode{}\nwbegindocs{36}%

Although we cannot guarantee that \impl{thread\_count} is zero when no threads are alive, we know there must be one call to \impl{dec\_thread\_count}, where \impl{thread\_count} reaches zero. Now if there are any tasks left, it may be the case that \impl{add\_task} already checked the value of \impl{thread\_count} before it got zero. Therefore we create a new thread. Note that this may lead to creating more threads than we actually need. However this may only affect the performance, not the correctness of the execution. We should nevertheless make this situation as unlikely as possible by increasing \impl{thread\_count} again as fast as possible after checking that it is zero. Therefore \impl{launch\_thread}, where we increase the thread counter, is set \impl{inline}. Also we refrain from checking against \impl{MAX\_THREADS}, since having this many threads is quite unlikely (although not impossible) in this situation.\\

If \impl{task\_count} is zero, we do not want to create a new thread (else we would end up in an infinte loop of creating and destroying new threads). However, this additional condition makes our safety check vulnerable. Therefore it is important that \impl{task\_count} never underestimates the number of tasks. Because of that the number of tasks inside the queue must be zero and we can safely assume that, if a new task is added, the method \impl{add\_task} will start a new thread.\\

For using \impl{std::thread}, we need to include the respective header
\nwenddocs{}\nwbegincode{37}\sublabel{NWwxjkf-2sbsH1-3}\nwmargintag{{\nwtagstyle{}\subpageref{NWwxjkf-2sbsH1-3}}}\moddef{tm includes~{\nwtagstyle{}\subpageref{NWwxjkf-2sbsH1-1}}}\plusendmoddef\nwstartdeflinemarkup\nwusesondefline{\\{NWwxjkf-47sd8i-1}}\nwprevnextdefs{NWwxjkf-2sbsH1-2}{\relax}\nwenddeflinemarkup
#include <thread>
\nwused{\\{NWwxjkf-47sd8i-1}}\nwendcode{}\nwbegindocs{38}%

\paragraph{Initialization and destruction}
\ \\
Since we want to implement a singleton, constructors and destructors should not be accessible from the outside. So we declare the copy constructor as well as the default destructor private.
\nwenddocs{}\nwbegincode{39}\sublabel{NWwxjkf-2SHrAR-1}\nwmargintag{{\nwtagstyle{}\subpageref{NWwxjkf-2SHrAR-1}}}\moddef{tm constructors and destructors~{\nwtagstyle{}\subpageref{NWwxjkf-2SHrAR-1}}}\endmoddef\nwstartdeflinemarkup\nwusesondefline{\\{NWwxjkf-47sd8i-1}}\nwprevnextdefs{\relax}{NWwxjkf-2SHrAR-2}\nwenddeflinemarkup
Thread_manager(Thread_manager const&)\{\};
~Thread_manager()\{\};
\nwalsodefined{\\{NWwxjkf-2SHrAR-2}}\nwused{\\{NWwxjkf-47sd8i-1}}\nwendcode{}\nwbegindocs{40}%

In the constructor, the counters for tasks and threads are initially set to zero, whereas the priority queue will initially reserve space for up to 20 tasks. This space will be automatically extended by the queue, when it is exhausted.
\nwenddocs{}\nwbegincode{41}\sublabel{NWwxjkf-2SHrAR-2}\nwmargintag{{\nwtagstyle{}\subpageref{NWwxjkf-2SHrAR-2}}}\moddef{tm constructors and destructors~{\nwtagstyle{}\subpageref{NWwxjkf-2SHrAR-1}}}\plusendmoddef\nwstartdeflinemarkup\nwusesondefline{\\{NWwxjkf-47sd8i-1}}\nwprevnextdefs{NWwxjkf-2SHrAR-1}{\relax}\nwenddeflinemarkup
Thread_manager():tasks(20),thread_count(0),task_count(0)\{
  assert(tasks.is_lock_free());
  assert(thread_count.is_lock_free());
  assert(task_count.is_lock_free());
\}
\nwused{\\{NWwxjkf-47sd8i-1}}\nwendcode{}\nwbegindocs{42}%
We also check on construction, whether the implementation of the queue and the atomic variables is lock-free. These are only soft assertions, i.e., if these assertions fail it will not affect the correctness of the implementation. However, the performance may be drastically affected, which is why we decided to place these assertions. To be able to use \impl{assert}, we need to include its header.
\nwenddocs{}\nwbegincode{43}\sublabel{NWwxjkf-3yauSd-1}\nwmargintag{{\nwtagstyle{}\subpageref{NWwxjkf-3yauSd-1}}}\moddef{tm debug includes~{\nwtagstyle{}\subpageref{NWwxjkf-3yauSd-1}}}\endmoddef\nwstartdeflinemarkup\nwusesondefline{\\{NWwxjkf-47sd8i-1}}\nwenddeflinemarkup
#include <cassert>
\nwused{\\{NWwxjkf-47sd8i-1}}\nwendcode{}\nwbegindocs{44}%

Finally we need to initialize the \impl{instance} pointer. Since it is static, it cannot be initialized in the header. So we need to create a source file, where we initialize \impl{instance} to be a null pointer.

\nwenddocs{}\nwbegincode{45}\sublabel{NWwxjkf-47smee-1}\nwmargintag{{\nwtagstyle{}\subpageref{NWwxjkf-47smee-1}}}\moddef{Thread\_manager.cpp~{\nwtagstyle{}\subpageref{NWwxjkf-47smee-1}}}\endmoddef\nwstartdeflinemarkup\nwenddeflinemarkup
#ifndef RA_THREAD_MANAGER_CPP
#define RA_THREAD_MANAGER_CPP

#ifdef RA_CXX11

#include "Thread_manager.hpp"

Thread_manager* Thread_manager::instance = nullptr;

#endif//RA_CXX11
#endif//RA_THREAD_MANAGER_CPP
\nwnotused{Thread\_manager.cpp}\nwendcode{}\nwbegindocs{46}%

\section{DAG Node Implementation}
\label{sss:threadapplication}

The goal of the following section is to parallelize the evaluation of the expression dag based on the evaluation model of the \impl{sdag\_node}. We will focus on parallelizing the accuracy-driven computation, since the main part of the complexity in the form of bigfloat operations lies there.

\paragraph{Outline}
\ \\
In 2015 Mörig and Schirra suggested using a topological evaluation order instead of the recursive procedure used in \impl{sdag\_node} for the accuracy-driven evaluation~\cite{moerig15macis}. This does not only fix some performance issues, it also enables us to parallelize the evaluation of nodes that are not dependent on each other, i.e., those nodes that are not connected through a directed path. We therefore assume that the evaluation is done in topological order. However we do only comment on the parts of the evaluation that must be adjusted for multithreaded evaluation. The remaining code can be found in the appendix.

\subsection{Ensuring Multithread-safety}
\label{ssc:multisafe}

Before we can parallelize the evaluation of independent nodes, we have to make sure that no data conflicts occur between such nodes. In particular we must remove or guard each write access to a child node during recomputation. Otherwise multiple, independent parents of the same node could get in conflict with each other.

There are two situations in which a non-read-only access to a child node can occur. These are
\begin{enumerate}
  \item Reference handling
  \item Computation of an algebraic degree bound for the node
\end{enumerate}
In the first case a parent node may delete its reference to a child node during computation, when it is converted to a bigfloat. We will handle this by ensuring that no access to the reference counter can corrupt it.

The second case is more troublesome. After each evaluation a separation bound for the current node is computed to decide whether the node evaluates to zero. For this we need to know an upper bound on the algebraic degree of the current node, which also may change (get smaller) during the computation if roots cancel out or can be computed exactly. 

In previous dag node variants the algebraic degree bound of an expression gets recomputed top-down recursively \emph{every time} a node gets recomputed. During this process, a temporary status member variable is used to indicate whether the node was already visited. This variable may get corrupted when multiple parents try to recompute their algebraic degree bound simultaneously. We solve this conflict by replacing the top-down computation by an bottom-up approach, in which each parent only needs read-only access to child node data.
In addition to making it multithread safe, this also replaces the previously quadratic runtime of the degree computation by an almost linear one\footnote{Due to the used data structures, the runtime of the implementation will not be linear in theory, but for almost all practical purposes.}.

\paragraph{Reference handling}
\ \\
As mentioned before, we need to ensure that simultaneous write access to the reference counter is handled correctly. This can be done by making the counter atomic.
\nwenddocs{}\nwbegincode{47}\sublabel{NWwxjkf-2obXTf-1}\nwmargintag{{\nwtagstyle{}\subpageref{NWwxjkf-2obXTf-1}}}\moddef{reference handling~{\nwtagstyle{}\subpageref{NWwxjkf-2obXTf-1}}}\endmoddef\nwstartdeflinemarkup\nwusesondefline{\\{NWwxjkf-qfj8R-1}}\nwprevnextdefs{\relax}{NWwxjkf-2obXTf-2}\nwenddeflinemarkup
std::atomic<unsigned int> count;
\nwalsodefined{\\{NWwxjkf-2obXTf-2}}\nwused{\\{NWwxjkf-qfj8R-1}}\nwendcode{}\nwbegindocs{48}%
By this each increment and decrement operation, \emph{including the return of its value}, will be executed atomically. Since the used access methods consist of only one such operation each and \impl{ref\_minus} returns by value, we can use the same code as before for them.

\nwenddocs{}\nwbegincode{49}\sublabel{NWwxjkf-2obXTf-2}\nwmargintag{{\nwtagstyle{}\subpageref{NWwxjkf-2obXTf-2}}}\moddef{reference handling~{\nwtagstyle{}\subpageref{NWwxjkf-2obXTf-1}}}\plusendmoddef\nwstartdeflinemarkup\nwusesondefline{\\{NWwxjkf-qfj8R-1}}\nwprevnextdefs{NWwxjkf-2obXTf-1}{\relax}\nwenddeflinemarkup
inline void ref_plus()\{ ++count; \}
inline unsigned int ref_minus()\{ return --count; \}
\nwused{\\{NWwxjkf-qfj8R-1}}\nwendcode{}\nwbegindocs{50}%

Making the counter atomic instead of using a lock mechanism is quite convenient and saves some overhead compared to maintaining a mutex. However, since most calls to \impl{ref\_plus} and \impl{ref\_minus} do not need to be multithread safe, it might actually be more efficient to use a lock in the few situations, when a node is converted to a bigfloat during accuracy-driven computation.

We chose the atomic variant over the use of mutexes, because in the current implementation we managed to avoid maintaining a mutex for each node and the overhead of atomic counters over standard counters is negligible compared to the rest of the computation. If in future implementations a need for mutexes emerges independently from reference handling, one may consider replacing the usage of atomic by mutexes.

\paragraph{Algebraic degree bound computation}
\ \\
In the following paragraphs we will replace the top-down approach in degree computation by a bottom-up approach. An upper bound for the algebraic degree of an expression in the form of an expression dag can be found by multiplying the degree of all of its nodes containing a root operation. In a tree, this can be done by recursively multiplying the degree bounds of a node's children. However, in a dag this may result in factoring in nodes with multiple references more than once and consequently may lead to a strong overestimation of the algebraic degree. So we must take this into consideration.\\

We develop our algorithm around the premise that usually the number of nodes, which influence the algebraic degree of an expression, is fairly small. If this would not be the case, then the separation bound we get would get very large, probably too large for any practical application, and therefore the cost associated with the evaluation would outweigh any cost associated with the degree bound computation.

Based on this premise, we (temporarily) store the algebraic degree bound of a node during evaluation by splitting it up into a variable called \impl{unique\_algebraic\_degree} and a collection of descendants with degree greater than one.
\nwenddocs{}\nwbegincode{51}\sublabel{NWwxjkf-3R0KsL-1}\nwmargintag{{\nwtagstyle{}\subpageref{NWwxjkf-3R0KsL-1}}}\moddef{algebraic degree computation members~{\nwtagstyle{}\subpageref{NWwxjkf-3R0KsL-1}}}\endmoddef\nwstartdeflinemarkup\nwusesondefline{\\{NWwxjkf-qfj8R-1}}\nwprevnextdefs{\relax}{NWwxjkf-3R0KsL-2}\nwenddeflinemarkup
Exponent unique_algebraic_degree;
std::vector<SelfType const*> vardeg_nodes;
\nwalsodefined{\\{NWwxjkf-3R0KsL-2}\\{NWwxjkf-3R0KsL-3}}\nwused{\\{NWwxjkf-qfj8R-1}}\nwendcode{}\nwbegindocs{52}%
In \impl{unique\_algebraic\_degree} the part of the algebraic degree bound will be stored, which is unique to this node in the sense that it does not contain the degree bounds of any descendants which may be used again by one of its parents. Every descendant with a degree greater than one that may be used again is stored in \impl{vardeg\_nodes}.

The operations we will need for our collection are union and find, both of which are inefficient in an array-like data structure like a vector. Nevertheless we still use a vector, since we expect the number of elements inside the vector to be very small, since these elements must have both
\begin{itemize}
  \item A \impl{unique\_degree} larger than one and
  \item More than one direct parent.
\end{itemize}
So presumably the number of elements in the vector will be almost always smaller than ten, in many cases even zero or one. For collections of this size the overhead of more complex data structures is not worth the performance gain, if there even is any gain at all.\\

The bound for the algebraic degree will then be computed by multiplying the unique parts of the node itself and of all nodes in \impl{vardeg\_nodes}.
\nwenddocs{}\nwbegincode{53}\sublabel{NWwxjkf-4cPL07-1}\nwmargintag{{\nwtagstyle{}\subpageref{NWwxjkf-4cPL07-1}}}\moddef{separation bound computation compute degree~{\nwtagstyle{}\subpageref{NWwxjkf-4cPL07-1}}}\endmoddef\nwstartdeflinemarkup\nwusesondefline{\\{NWwxjkf-3rifsU-1}}\nwprevnextdefs{\relax}{NWwxjkf-4cPL07-2}\nwenddeflinemarkup
Exponent get_algebraic_degree() const\{

  assert(unique_algebraic_degree >= 1);

  Exponent algebraic_degree = unique_algebraic_degree;
  for (auto node : vardeg_nodes) \{
    algebraic_degree *= node->get_algebraic_degree();
  \}

  return algebraic_degree;
\}
\nwalsodefined{\\{NWwxjkf-4cPL07-2}\\{NWwxjkf-4cPL07-3}\\{NWwxjkf-4cPL07-4}}\nwused{\\{NWwxjkf-3rifsU-1}}\nwendcode{}\nwbegindocs{54}%

To identify which nodes have multiple parents, we use a temporary variable, which will be set during the preprocessing, i.e., the topological sorting.
\nwenddocs{}\nwbegincode{55}\sublabel{NWwxjkf-3R0KsL-2}\nwmargintag{{\nwtagstyle{}\subpageref{NWwxjkf-3R0KsL-2}}}\moddef{algebraic degree computation members~{\nwtagstyle{}\subpageref{NWwxjkf-3R0KsL-1}}}\plusendmoddef\nwstartdeflinemarkup\nwusesondefline{\\{NWwxjkf-qfj8R-1}}\nwprevnextdefs{NWwxjkf-3R0KsL-1}{NWwxjkf-3R0KsL-3}\nwenddeflinemarkup
bool has_multiple_references;
\nwused{\\{NWwxjkf-qfj8R-1}}\nwendcode{}\nwbegindocs{56}%

During preprocessing we will also reset all temporary parameters. The \impl{unique\_algebraic\_degree} will be set to the degree of the node, which is $d$ for the $d$-th root and $1$ otherwise.
\nwenddocs{}\nwbegincode{57}\sublabel{NWwxjkf-3R0KsL-3}\nwmargintag{{\nwtagstyle{}\subpageref{NWwxjkf-3R0KsL-3}}}\moddef{algebraic degree computation members~{\nwtagstyle{}\subpageref{NWwxjkf-3R0KsL-1}}}\plusendmoddef\nwstartdeflinemarkup\nwusesondefline{\\{NWwxjkf-qfj8R-1}}\nwprevnextdefs{NWwxjkf-3R0KsL-2}{\relax}\nwenddeflinemarkup
void reset_algebraic_degree_parameters() \{
  unique_algebraic_degree = degree();
  vardeg_nodes.clear();
  has_multiple_references = false;
\}
\nwused{\\{NWwxjkf-qfj8R-1}}\nwendcode{}\nwbegindocs{58}%

Now during postprocessing of the computation step we compute the new degree bound for the node by factoring in its child nodes.
\nwenddocs{}\nwbegincode{59}\sublabel{NWwxjkf-4cPL07-2}\nwmargintag{{\nwtagstyle{}\subpageref{NWwxjkf-4cPL07-2}}}\moddef{separation bound computation compute degree~{\nwtagstyle{}\subpageref{NWwxjkf-4cPL07-1}}}\plusendmoddef\nwstartdeflinemarkup\nwusesondefline{\\{NWwxjkf-3rifsU-1}}\nwprevnextdefs{NWwxjkf-4cPL07-1}{NWwxjkf-4cPL07-3}\nwenddeflinemarkup
void compute_degree_bound()\{
  assert(unique_algebraic_degree == degree());
  alg_degree_factor_in(X);
  alg_degree_factor_in(Y);
\}
\nwused{\\{NWwxjkf-3rifsU-1}}\nwendcode{}\nwbegindocs{60}%

If a child exists, we need to decide, whether it has multiple parents and needs to be stored in \impl{vardeg\_nodes}. Either way we must merge the \impl{vardeg\_nodes} of the child with our own collection (which will be initially empty). 
\nwenddocs{}\nwbegincode{61}\sublabel{NWwxjkf-4cPL07-3}\nwmargintag{{\nwtagstyle{}\subpageref{NWwxjkf-4cPL07-3}}}\moddef{separation bound computation compute degree~{\nwtagstyle{}\subpageref{NWwxjkf-4cPL07-1}}}\plusendmoddef\nwstartdeflinemarkup\nwusesondefline{\\{NWwxjkf-3rifsU-1}}\nwprevnextdefs{NWwxjkf-4cPL07-2}{NWwxjkf-4cPL07-4}\nwenddeflinemarkup
void alg_degree_factor_in(SelfType const* child)\{
  if (child) \{
    \LA{}factor in child unique degree~{\nwtagstyle{}\subpageref{NWwxjkf-3BswN0-1}}\RA{}
    \LA{}merge child vardeg nodes~{\nwtagstyle{}\subpageref{NWwxjkf-1hEUs5-1}}\RA{}
  \}
\}
\nwused{\\{NWwxjkf-3rifsU-1}}\nwendcode{}\nwbegindocs{62}%

The only situation in which we must consider a child's unique degree at all is if it is greater than one. If the child node has multiple parents, we must make sure that it is not counted more than once and therefore add it to our collection, else it cannot occur anywhere else and therefore we can include it in the node's unique degree.
\nwenddocs{}\nwbegincode{63}\sublabel{NWwxjkf-3BswN0-1}\nwmargintag{{\nwtagstyle{}\subpageref{NWwxjkf-3BswN0-1}}}\moddef{factor in child unique degree~{\nwtagstyle{}\subpageref{NWwxjkf-3BswN0-1}}}\endmoddef\nwstartdeflinemarkup\nwusesondefline{\\{NWwxjkf-4cPL07-3}}\nwenddeflinemarkup
if (child->unique_algebraic_degree > 1) \{
  if (child->has_multiple_references) \{
    add_var_deg_node(child);
  \} else \{
    unique_algebraic_degree *= child->unique_algebraic_degree;
  \}
\}
\nwused{\\{NWwxjkf-4cPL07-3}}\nwendcode{}\nwbegindocs{64}%

If the child has any \impl{vardeg\_nodes} itself, we must merge them with our own, which is, we effectively merge the sets of \impl{vardeg\_nodes} of our left and right child. By ensuring that no node appears twice in our collection, we also ensure that none of those degrees contributes to our final algebraic degree bound more than once.
\nwenddocs{}\nwbegincode{65}\sublabel{NWwxjkf-1hEUs5-1}\nwmargintag{{\nwtagstyle{}\subpageref{NWwxjkf-1hEUs5-1}}}\moddef{merge child vardeg nodes~{\nwtagstyle{}\subpageref{NWwxjkf-1hEUs5-1}}}\endmoddef\nwstartdeflinemarkup\nwusesondefline{\\{NWwxjkf-4cPL07-3}}\nwenddeflinemarkup
for (auto node : child->vardeg_nodes) \{
  add_var_deg_node(node);  
\}
\nwused{\\{NWwxjkf-4cPL07-3}}\nwendcode{}\nwbegindocs{66}%
Since we do not have a set union operation in a vector, we must iterate through all nodes in the first collection and insert them into the second one. So we search for each node in the parent's vector and, if it is not already there, we insert it at the end.
\nwenddocs{}\nwbegincode{67}\sublabel{NWwxjkf-4cPL07-4}\nwmargintag{{\nwtagstyle{}\subpageref{NWwxjkf-4cPL07-4}}}\moddef{separation bound computation compute degree~{\nwtagstyle{}\subpageref{NWwxjkf-4cPL07-1}}}\plusendmoddef\nwstartdeflinemarkup\nwusesondefline{\\{NWwxjkf-3rifsU-1}}\nwprevnextdefs{NWwxjkf-4cPL07-3}{\relax}\nwenddeflinemarkup
inline void add_var_deg_node(SelfType const* node) \{
  if (std::find(vardeg_nodes.begin(),vardeg_nodes.end(),node) == vardeg_nodes.end()) \{
    vardeg_nodes.push_back(node);
  \}
\}
\nwused{\\{NWwxjkf-3rifsU-1}}\nwendcode{}\nwbegindocs{68}%

\paragraph{Separation bound computation}
\ \\
We need the algebraic degree bound for the computation of a separation bound. To be able to enforce const correctness, which is very useful in a multithreaded environment, we first introduce a new getter method for the separation bound object. 
\nwenddocs{}\nwbegincode{69}\sublabel{NWwxjkf-2Ingyi-1}\nwmargintag{{\nwtagstyle{}\subpageref{NWwxjkf-2Ingyi-1}}}\moddef{separation bound getter and setter~{\nwtagstyle{}\subpageref{NWwxjkf-2Ingyi-1}}}\endmoddef\nwstartdeflinemarkup\nwusesondefline{\\{NWwxjkf-1qdLgq-1}}\nwprevnextdefs{\relax}{NWwxjkf-2Ingyi-2}\nwenddeflinemarkup
inline SeparationBound const& get_sep_bd() const \{ return node_data->sep_bd; \}
\nwalsodefined{\\{NWwxjkf-2Ingyi-2}}\nwused{\\{NWwxjkf-1qdLgq-1}}\nwendcode{}\nwbegindocs{70}%
Then we use this getter and the newly defined method \impl{get\_algebraic\_degree} to call the separation bound computation.
\nwenddocs{}\nwbegincode{71}\sublabel{NWwxjkf-3PC6VY-1}\nwmargintag{{\nwtagstyle{}\subpageref{NWwxjkf-3PC6VY-1}}}\moddef{separation bound computation wrapper~{\nwtagstyle{}\subpageref{NWwxjkf-3PC6VY-1}}}\endmoddef\nwstartdeflinemarkup\nwusesondefline{\\{NWwxjkf-3rifsU-1}}\nwenddeflinemarkup
inline Exponent sep_bound() const\{
  return get_sep_bd().bound(get_algebraic_degree());
\}
\nwused{\\{NWwxjkf-3rifsU-1}}\nwendcode{}\nwbegindocs{72}%
It is important that \impl{compute\_degree\_bound} is called before \impl{sep\_bound}, as well as that all children of the current node have already been recomputed. Since there is no way of detecting whether the computation has already been executed, in some cases this leads to errors, more precisely to underestimations of the algebraic degree, that are hard to find.

We summarize the methods in a separation bound computation chunk.
\nwenddocs{}\nwbegincode{73}\sublabel{NWwxjkf-3rifsU-1}\nwmargintag{{\nwtagstyle{}\subpageref{NWwxjkf-3rifsU-1}}}\moddef{separation bound computation~{\nwtagstyle{}\subpageref{NWwxjkf-3rifsU-1}}}\endmoddef\nwstartdeflinemarkup\nwusesondefline{\\{NWwxjkf-qfj8R-1}}\nwenddeflinemarkup
\LA{}separation bound computation wrapper~{\nwtagstyle{}\subpageref{NWwxjkf-3PC6VY-1}}\RA{}
\LA{}separation bound computation compute degree~{\nwtagstyle{}\subpageref{NWwxjkf-4cPL07-1}}\RA{}
\LA{}separation bound concrete computation~{\nwtagstyle{}\subpageref{NWwxjkf-4Litu7-1}}\RA{}
\nwused{\\{NWwxjkf-qfj8R-1}}\nwendcode{}\nwbegindocs{74}%

The separation bound is used in the method \impl{fixup\_post\_recompute} directly after the recomputation of a node's approximation to decide whether the value of the node is zero (see also the implementation of \impl{recompute\_op} in Section~\ref{ssc:evaluation}). We adjust this method to account for the initialization of our new algebraic degree bound computation parameters.

\nwenddocs{}\nwbegincode{75}\sublabel{NWwxjkf-3j4HBX-1}\nwmargintag{{\nwtagstyle{}\subpageref{NWwxjkf-3j4HBX-1}}}\moddef{adc fixup~{\nwtagstyle{}\subpageref{NWwxjkf-3j4HBX-1}}}\endmoddef\nwstartdeflinemarkup\nwusesondefline{\\{NWwxjkf-hyGrK-1}}\nwenddeflinemarkup
void fixup_post_recompute()\{
   
  if(exact())\{
    convert_to_bigfloat();
    \LA{}reset parameters~{\nwtagstyle{}\subpageref{NWwxjkf-26Om8R-1}}\RA{}
  \} else \{
    \LA{}check if node is exactly zero~{\nwtagstyle{}\subpageref{NWwxjkf-4Carso-1}}\RA{}
  \}

  \LA{}try to improve filter policy~{\nwtagstyle{}\subpageref{NWwxjkf-KEXze-1}}\RA{}
\}
\nwused{\\{NWwxjkf-hyGrK-1}}\nwendcode{}\nwbegindocs{76}%

Whenever the node is converted to a bigfloat during the computation, we have to reset our computation parameters. This can happen whenever an exact value is detected, i.e., when the approximation is exact or when the node can be identified as exactly zero. As before, we also reset the separation bound.
\nwenddocs{}\nwbegincode{77}\sublabel{NWwxjkf-26Om8R-1}\nwmargintag{{\nwtagstyle{}\subpageref{NWwxjkf-26Om8R-1}}}\moddef{reset parameters~{\nwtagstyle{}\subpageref{NWwxjkf-26Om8R-1}}}\endmoddef\nwstartdeflinemarkup\nwusesondefline{\\{NWwxjkf-3j4HBX-1}\\{NWwxjkf-4Carso-1}}\nwenddeflinemarkup
init_sep_bd();
reset_algebraic_degree_parameters();
\nwused{\\{NWwxjkf-3j4HBX-1}\\{NWwxjkf-4Carso-1}}\nwendcode{}\nwbegindocs{78}%

If the approximation is not exact, we check for the separation bound. Before that, we have to compute the bound for the algebraic degree. Note that, since this is not done recursively, we must ensure that the \impl{compute\_degree\_bound} method is called in every (non-exact) node, irrespective of whether the separation bound actually gets computed or not.

\nwenddocs{}\nwbegincode{79}\sublabel{NWwxjkf-4Carso-1}\nwmargintag{{\nwtagstyle{}\subpageref{NWwxjkf-4Carso-1}}}\moddef{check if node is exactly zero~{\nwtagstyle{}\subpageref{NWwxjkf-4Carso-1}}}\endmoddef\nwstartdeflinemarkup\nwusesondefline{\\{NWwxjkf-3j4HBX-1}}\nwenddeflinemarkup
init_sep_bd();
compute_degree_bound();

if( is_zero() || 
    floor_log2_approx() <= ceil_log2_error() )\{
  Exponent q = get_requested_error();
        
  if(q + 1 < sep_bound())\{
    zeroize();
    convert_to_bigfloat();
    \LA{}reset parameters~{\nwtagstyle{}\subpageref{NWwxjkf-26Om8R-1}}\RA{}
  \}
\}
\nwused{\\{NWwxjkf-3j4HBX-1}}\nwendcode{}\nwbegindocs{80}%

If the node can be evaluated to zero and therefore converted to a bigfloat, we again have to reset our computation parameters.

\subsection{Parallelizing the Evaluation}
\label{ssc:evaluation}

After making the node class multithread-safe, we can finally parallelize the evaluation. For this we will use the thread manager as described in Section~\ref{sss:threadmanagement}.

\paragraph{Parallel Recomputation}
\ \\
We will only focus on the recomputation of the approximation. In particular, we will not try to parallelize the topological sorting or the error propagation.
\nwenddocs{}\nwbegincode{81}\sublabel{NWwxjkf-1RZNmO-1}\nwmargintag{{\nwtagstyle{}\subpageref{NWwxjkf-1RZNmO-1}}}\moddef{adc wrapper~{\nwtagstyle{}\subpageref{NWwxjkf-1RZNmO-1}}}\endmoddef\nwstartdeflinemarkup\nwusesondefline{\\{NWwxjkf-hyGrK-1}}\nwprevnextdefs{\relax}{NWwxjkf-1RZNmO-2}\nwenddeflinemarkup
void guarantee_bound_two_to(const Exponent& q)\{
  if(exact() || ceil_log2_error() <= q) return;
  
  std::vector<SelfType*> order;
  topsort_visit(this,order);

  requested_error() = q;
  
  \LA{}propagate error topologically descending~{\nwtagstyle{}\subpageref{NWwxjkf-6vwGL-1}}\RA{}
  \LA{}recompute approximation thread based~{\nwtagstyle{}\subpageref{NWwxjkf-3jZbw2-1}}\RA{}
\}
\nwalsodefined{\\{NWwxjkf-1RZNmO-2}}\nwused{\\{NWwxjkf-hyGrK-1}}\nwendcode{}\nwbegindocs{82}%
However, we still have to adjust the implementation of the topological sorting to fit our needs (see `Preprocessing'). Our strategy for recomputation is as follows
\begin{enumerate}
  \item Recompute all leaf nodes and notify their parents when the computation is finished.
  \item Start the computation of an inner node as soon as all its children have sent a notification.
  \item When the root node has been recomputed, wake up the main thread and continue its execution.
\end{enumerate}

So in this strategy, the responsibility of starting the recomputation of inner nodes can be shifted from the main thread to the nodes and is therefore decentralized. This has several advantages:
\begin{enumerate}
  \item There is no need for an expensive communication strategy between the main thread and the subthreads, except for the notification by the root node. 
  \item We can start new tasks on demand and therefore whenever a task is started, we can guarantee that its requirements are satisfied. Consequently, we never have to wait for any requirement during execution. 
  \item This allows us to use a notification mechanism, which can be executed lock-free\footnote{It means without the need for semaphores or other locking mechanisms.}, except for the communication between root node and main thread.
  \item We do not depend on the thread manager to detect, when the computation is finished.
\end{enumerate}

On the downside, each node needs to know its parents in the current part of a (possibly larger) dag, in which the recomputation is executed. Therefore we need to assign all current parents to a node during preparation, i.e., during the topological sorting. We save them in a local vector.
\nwenddocs{}\nwbegincode{83}\sublabel{NWwxjkf-XH50e-1}\nwmargintag{{\nwtagstyle{}\subpageref{NWwxjkf-XH50e-1}}}\moddef{parent handling~{\nwtagstyle{}\subpageref{NWwxjkf-XH50e-1}}}\endmoddef\nwstartdeflinemarkup\nwusesondefline{\\{NWwxjkf-qfj8R-1}}\nwprevnextdefs{\relax}{NWwxjkf-XH50e-2}\nwenddeflinemarkup
std::vector<SelfType*> parents;
\nwalsodefined{\\{NWwxjkf-XH50e-2}\\{NWwxjkf-XH50e-3}\\{NWwxjkf-XH50e-4}}\nwused{\\{NWwxjkf-qfj8R-1}}\nwendcode{}\nwbegindocs{84}%

Furthermore each parent node needs to know how many dependencies it has, i.e., how many notifications are necessary until its recomputation can be started. This number may be different from the number of children as we will see later, since not all nodes need recomputation.
\nwenddocs{}\nwbegincode{85}\sublabel{NWwxjkf-XH50e-2}\nwmargintag{{\nwtagstyle{}\subpageref{NWwxjkf-XH50e-2}}}\moddef{parent handling~{\nwtagstyle{}\subpageref{NWwxjkf-XH50e-1}}}\plusendmoddef\nwstartdeflinemarkup\nwusesondefline{\\{NWwxjkf-qfj8R-1}}\nwprevnextdefs{NWwxjkf-XH50e-1}{NWwxjkf-XH50e-3}\nwenddeflinemarkup
std::atomic<int> dependency_count;
\nwused{\\{NWwxjkf-qfj8R-1}}\nwendcode{}\nwbegindocs{86}%

It is very important that \impl{dependency\_count} is atomic, since its children may simultaneously try to notify their parent and therefore decrease its dependency counter. This issue will become apparent soon.

To ensure that the number of notifying children fits the number of dependencies, we will put their initialization together in one method. 
\nwenddocs{}\nwbegincode{87}\sublabel{NWwxjkf-XH50e-3}\nwmargintag{{\nwtagstyle{}\subpageref{NWwxjkf-XH50e-3}}}\moddef{parent handling~{\nwtagstyle{}\subpageref{NWwxjkf-XH50e-1}}}\plusendmoddef\nwstartdeflinemarkup\nwusesondefline{\\{NWwxjkf-qfj8R-1}}\nwprevnextdefs{NWwxjkf-XH50e-2}{NWwxjkf-XH50e-4}\nwenddeflinemarkup
inline void register_parent(SelfType* p) \{ 
  parents.push_back(p);
  ++(p->dependency_count);
\}
\nwused{\\{NWwxjkf-qfj8R-1}}\nwendcode{}\nwbegindocs{88}%

We will now get to the core method that decides when to launch a new task.
\nwenddocs{}\nwbegincode{89}\sublabel{NWwxjkf-XH50e-4}\nwmargintag{{\nwtagstyle{}\subpageref{NWwxjkf-XH50e-4}}}\moddef{parent handling~{\nwtagstyle{}\subpageref{NWwxjkf-XH50e-1}}}\plusendmoddef\nwstartdeflinemarkup\nwusesondefline{\\{NWwxjkf-qfj8R-1}}\nwprevnextdefs{NWwxjkf-XH50e-3}{\relax}\nwenddeflinemarkup
void try_recomputation () \{
  if (dependency_count-- == 0)  \{
    \LA{}start new task~{\nwtagstyle{}\subpageref{NWwxjkf-2itq2c-1}}\RA{}
  \}
\}
\nwused{\\{NWwxjkf-qfj8R-1}}\nwendcode{}\nwbegindocs{90}%
Calling the method \impl{try\_recomputation} equates to notifying the node that one of its dependencies has been fulfilled. If there are no dependencies left, 
its recomputation task will be started. We decrease the dependency counter \emph{after} checking whether it is zero, since this method will be called once from the main thread for each node, to start those that have no dependencies. So for each node this method will be called exactly \impl{dependency\_count+1} times and the task will be started during the last call\footnote{This refers to the call, which decreased the counter last, not necessarily the call, which has been started last}, although we do not know which call will be the last.\\

It is necessary for the decrease operation and the check to happen atomically, i.e., for \impl{dependency\_count} to be atomic. If the counter would be decreased in a separate operation, e.g. in an else block, it would be possible that the new task is never started. This happens, if the following sequence of events occurs:
\begin{enumerate}
  \item child $a$ and child $b$ want to notify their parent node with \impl{dependency\_count} $1$
  \item child $a$ checks \impl{dependency\_count}, which is $1$
  \item child $b$ checks \impl{dependency\_count}, which is $1$
  \item child $a$ decreases \impl{dependency\_count} to $0$
  \item child $b$ decreases \impl{dependency\_count} to $-1$
\end{enumerate}
To make things worse, accessing any member variable after the check may lead the program to crash or, even worse, produce wrong results, since the execution may have already stopped at this point or a new computation may have started.\\

It should also be noted that the atomic keyword guarantees that the return value of a pre- or post-decrement operation is given by value. So even if another decrement occurs in between the previous post-decrement and the evaluation of the equals operator, it will not affect the result of the evaluation\footnote{Admittedly, return by reference should only be possible to occur for pre-decrement anyway.}.\\

We start a new task by binding the node to the \impl{recompute\_op} function (which we describe below) and passing it to the \impl{Thread\_manager}.
\nwenddocs{}\nwbegincode{91}\sublabel{NWwxjkf-2itq2c-1}\nwmargintag{{\nwtagstyle{}\subpageref{NWwxjkf-2itq2c-1}}}\moddef{start new task~{\nwtagstyle{}\subpageref{NWwxjkf-2itq2c-1}}}\endmoddef\nwstartdeflinemarkup\nwusesondefline{\\{NWwxjkf-XH50e-4}}\nwenddeflinemarkup
Thread_manager::get_instance()->add_task(
  new std::function<void()>(std::bind(recompute_op,this))
);
\nwused{\\{NWwxjkf-XH50e-4}}\nwendcode{}\nwbegindocs{92}%

As mentioned before, we try to start the computation of each node once inside the main thread.
\nwenddocs{}\nwbegincode{93}\sublabel{NWwxjkf-3jZbw2-1}\nwmargintag{{\nwtagstyle{}\subpageref{NWwxjkf-3jZbw2-1}}}\moddef{recompute approximation thread based~{\nwtagstyle{}\subpageref{NWwxjkf-3jZbw2-1}}}\endmoddef\nwstartdeflinemarkup\nwusesondefline{\\{NWwxjkf-1RZNmO-1}}\nwenddeflinemarkup
\LA{}prepare synchronization~{\nwtagstyle{}\subpageref{NWwxjkf-2y2CB9-1}}\RA{}

for(auto node : order) \{
  node->try_recomputation();
\}
\LA{}synchronize threads~{\nwtagstyle{}\subpageref{NWwxjkf-10iWX4-1}}\RA{}
\nwused{\\{NWwxjkf-1RZNmO-1}}\nwendcode{}\nwbegindocs{94}%
This will be the starting point for each node that does not have any dependencies. It can also be potentially the starting point of nodes, whose dependencies were resolved before the main loop reached them. After notifying each node once, we will pause the execution until all tasks are finished. This will be described in more detail in the paragraph `Synchronization'.\\

The \impl{recompute\_op} method takes a node and recomputes its approximation such that the error will be smaller than the requested error, assuming that the child nodes have already been recomputed with appropriate error bounds. Since this method will be passed to the thread manager, it must be static.

The recomputation itself happens in the same way as without parallelization. However, after the node is recomputed, it must now send a notification to its parents or, if it is the root node, to the main thread. Also note that we already slightly modified the fixup method in Section~\ref{ssc:multisafe} to implement the new algebraic degree bound computation strategy.
\nwenddocs{}\nwbegincode{95}\sublabel{NWwxjkf-1RZNmO-2}\nwmargintag{{\nwtagstyle{}\subpageref{NWwxjkf-1RZNmO-2}}}\moddef{adc wrapper~{\nwtagstyle{}\subpageref{NWwxjkf-1RZNmO-1}}}\plusendmoddef\nwstartdeflinemarkup\nwusesondefline{\\{NWwxjkf-hyGrK-1}}\nwprevnextdefs{NWwxjkf-1RZNmO-1}{\relax}\nwenddeflinemarkup
static void recompute_op(SelfType* node) \{

  assert(node->contype() != DOUBLE);
  assert(node->contype() != BIGFLOAT);

  switch(node->contype()) \{
    case NEGATION:       node->recompute_neg();  break;
    case ROOT:           node->recompute_root(); break;
    case ADDITION:       node->recompute_add();  break;
    case SUBTRACTION:    node->recompute_sub();  break;
    case MULTIPLICATION: node->recompute_mul();  break;
    case DIVISION:       node->recompute_div();  break;
  \}
        
  node->fixup_post_recompute();
  node->topsort_status() = CLEARED;

  \LA{}send notifications~{\nwtagstyle{}\subpageref{NWwxjkf-2vFwl8-1}}\RA{}
\}
\nwused{\\{NWwxjkf-hyGrK-1}}\nwendcode{}\nwbegindocs{96}%

The root node can be identified by checking, whether its \impl{parents} vector is empty. Since we only assign parents relevant to the recomputation, the root node will never have parents. On the contrary, each other node must have at least one parent, since otherwise it would not influence the value of the root node and therefore would not have been recomputed in the first place.
\nwenddocs{}\nwbegincode{97}\sublabel{NWwxjkf-2vFwl8-1}\nwmargintag{{\nwtagstyle{}\subpageref{NWwxjkf-2vFwl8-1}}}\moddef{send notifications~{\nwtagstyle{}\subpageref{NWwxjkf-2vFwl8-1}}}\endmoddef\nwstartdeflinemarkup\nwusesondefline{\\{NWwxjkf-1RZNmO-2}}\nwenddeflinemarkup
if (node->parents.empty()) \{
  \LA{}synchronize with main thread~{\nwtagstyle{}\subpageref{NWwxjkf-2U4UWm-1}}\RA{}
\} else \{
  \LA{}notify parents~{\nwtagstyle{}\subpageref{NWwxjkf-vBDTo-1}}\RA{}
\}
\nwused{\\{NWwxjkf-1RZNmO-2}}\nwendcode{}\nwbegindocs{98}%

If the node is not the root, we notify its parents that it is ready.
\nwenddocs{}\nwbegincode{99}\sublabel{NWwxjkf-vBDTo-1}\nwmargintag{{\nwtagstyle{}\subpageref{NWwxjkf-vBDTo-1}}}\moddef{notify parents~{\nwtagstyle{}\subpageref{NWwxjkf-vBDTo-1}}}\endmoddef\nwstartdeflinemarkup\nwusesondefline{\\{NWwxjkf-2vFwl8-1}}\nwenddeflinemarkup
for (auto p: node->parents) \{
  p->try_recomputation();
\}
\nwused{\\{NWwxjkf-2vFwl8-1}}\nwendcode{}\nwbegindocs{100}%

As before there cannot be any statement after the last call to \impl{try\_recomputation}, since from this point on we have no guarantees about the context anymore. The node could be deleted or another computation could have started after notifying the last parent.

\paragraph{Synchronization}
\ \\
While working on the tasks, we need our main thread to pause its execution. Unfortunately we cannot use this thread to compute tasks itself due to our decentralized strategy and the strict separation between the computation and the task manger. So we must wait until the root node notifies the main thread that its recomputation has finished. We do this by using a \emph{mutex} in combination with a \emph{condition variable}.
\nwenddocs{}\nwbegincode{101}\sublabel{NWwxjkf-3eVWfJ-1}\nwmargintag{{\nwtagstyle{}\subpageref{NWwxjkf-3eVWfJ-1}}}\moddef{threading members~{\nwtagstyle{}\subpageref{NWwxjkf-3eVWfJ-1}}}\endmoddef\nwstartdeflinemarkup\nwusesondefline{\\{NWwxjkf-qfj8R-1}}\nwprevnextdefs{\relax}{NWwxjkf-3eVWfJ-2}\nwenddeflinemarkup
static std::mutex sync_mutex;
static std::condition_variable sync_cv;
\nwalsodefined{\\{NWwxjkf-3eVWfJ-2}}\nwused{\\{NWwxjkf-qfj8R-1}}\nwendcode{}\nwbegindocs{102}%

Since we need only one of them, we make them static. We also need a boolean, indicating when the computation is finished.
\nwenddocs{}\nwbegincode{103}\sublabel{NWwxjkf-3eVWfJ-2}\nwmargintag{{\nwtagstyle{}\subpageref{NWwxjkf-3eVWfJ-2}}}\moddef{threading members~{\nwtagstyle{}\subpageref{NWwxjkf-3eVWfJ-1}}}\plusendmoddef\nwstartdeflinemarkup\nwusesondefline{\\{NWwxjkf-qfj8R-1}}\nwprevnextdefs{NWwxjkf-3eVWfJ-1}{\relax}\nwenddeflinemarkup
static bool computation_finished;
\nwused{\\{NWwxjkf-qfj8R-1}}\nwendcode{}\nwbegindocs{104}%
Since they are static template members, we must initialize all three variables outside the class declaration in the header file.
\nwenddocs{}\nwbegincode{105}\sublabel{NWwxjkf-2yBJ4v-1}\nwmargintag{{\nwtagstyle{}\subpageref{NWwxjkf-2yBJ4v-1}}}\moddef{static member definition~{\nwtagstyle{}\subpageref{NWwxjkf-2yBJ4v-1}}}\endmoddef\nwstartdeflinemarkup\nwusesondefline{\\{NWwxjkf-qfj8R-1}}\nwenddeflinemarkup
template <class P> std::mutex mtdag_node<P>::sync_mutex;
template <class P> std::condition_variable mtdag_node<P>::sync_cv;
template <class P> bool mtdag_node<P>::computation_finished = false;
\nwused{\\{NWwxjkf-qfj8R-1}}\nwendcode{}\nwbegindocs{106}%

While a mutex is simply a semaphore, which can be locked atomically and blocks all other lock attempts until the lock is freed, a condition variable functions as a permanent\footnote{Aside from so-called `spurious wakeups'.} block until it receives a notification from another thread and its condition is fulfilled. It is designed such that, while a thread is blocked through a condition variable, it consumes minimal time during context switches.\\

A condition variable always depends on an underlying mutex, which it will lock as soon as it wakes up to test for its condition. If the condition is not fulfilled yet, it will go to sleep again and unlock the mutex. This check also happens before waiting for the first notification, which is why the mutex needs to be locked before starting the waiting process.\\

So before starting any tasks, we set our condition, the boolean \impl{computation\_finished}, to \impl{false}.
\nwenddocs{}\nwbegincode{107}\sublabel{NWwxjkf-2y2CB9-1}\nwmargintag{{\nwtagstyle{}\subpageref{NWwxjkf-2y2CB9-1}}}\moddef{prepare synchronization~{\nwtagstyle{}\subpageref{NWwxjkf-2y2CB9-1}}}\endmoddef\nwstartdeflinemarkup\nwusesondefline{\\{NWwxjkf-3jZbw2-1}}\nwenddeflinemarkup
computation_finished = false;
\nwused{\\{NWwxjkf-3jZbw2-1}}\nwendcode{}\nwbegindocs{108}%

A \impl{unique\_lock} immediately locks the corresponding mutex on construction and unlocks it again either when its \impl{unlock} method is called, or when it gets destructed. After we tried to recompute each node once, we lock the synchronization mutex with a \impl{unique\_lock} and start the waiting process.
\nwenddocs{}\nwbegincode{109}\sublabel{NWwxjkf-10iWX4-1}\nwmargintag{{\nwtagstyle{}\subpageref{NWwxjkf-10iWX4-1}}}\moddef{synchronize threads~{\nwtagstyle{}\subpageref{NWwxjkf-10iWX4-1}}}\endmoddef\nwstartdeflinemarkup\nwusesondefline{\\{NWwxjkf-3jZbw2-1}}\nwenddeflinemarkup
std::unique_lock<std::mutex> lk(sync_mutex);
sync_cv.wait(lk,[this]()\{ return computation_finished; \});
\nwused{\\{NWwxjkf-3jZbw2-1}}\nwendcode{}\nwbegindocs{110}%

The condition variable lets the thread wait until it gets notified and its condition is \impl{true}. While it is waiting, \impl{sync\_cv} unlocks the mutex. If it wakes up, the mutex will be locked again. Note that a wakeup can also occur `spuriously', without any notification from other threads. Therefore it is important \emph{not} to set \impl{computation\_finished} to true before the root node has finished its computation.\\

Finally after recomputing the root, we set \impl{computation\_finished} to \impl{true} and notify the condition variable.
\nwenddocs{}\nwbegincode{111}\sublabel{NWwxjkf-2U4UWm-1}\nwmargintag{{\nwtagstyle{}\subpageref{NWwxjkf-2U4UWm-1}}}\moddef{synchronize with main thread~{\nwtagstyle{}\subpageref{NWwxjkf-2U4UWm-1}}}\endmoddef\nwstartdeflinemarkup\nwusesondefline{\\{NWwxjkf-2vFwl8-1}}\nwenddeflinemarkup
std::lock_guard<std::mutex> sync_lk(node->sync_mutex);

node->computation_finished = true;
node->sync_cv.notify_one();
\nwused{\\{NWwxjkf-2vFwl8-1}}\nwendcode{}\nwbegindocs{112}%
A \impl{lock\_guard} is a simple lock, which locks the mutex on creation and only unlocks on destruction. This makes it a bit more efficient than the previously used \impl{unique\_lock}.
We must lock \impl{sync\_mutex} before changing the condition and notifying \impl{sync\_cv} to prevent the following sequence of events from happening.
\begin{enumerate}
  \item \impl{computation\_finished} is set to false
  \item The computation gets finished
  \item \impl{sync\_cv} checks \impl{computation\_finished}
  \item \impl{computation\_finished} is set to true
  \item \impl{sync\_cv.notify\_one()} is called
  \item \impl{sync\_cv} goes to sleep
\end{enumerate}
By using the lock, the thread computing the root node must wait until \impl{sync\_cv} goes to sleep and unlocks the mutex, before notifying it\footnote{Except when it is called before the \impl{unique\_lock} is created. Then the main thread will never start waiting, since the condition is fulfilled from the beginning.}. Similarly, the condition variable must wait until the other threads \impl{lock\_guard} has been destroyed, before it can check its condition again.

\paragraph{Preprocessing}
\ \\
In this section, we have a look at the preprocessing that is necessary before each recomputation of the dag. Before starting the recomputation, a topological order must be determined and the temporary parameters must be initialized. We combine both steps in the method \impl{topsort\_visit}.
\nwenddocs{}\nwbegincode{113}\sublabel{NWwxjkf-Dyu9X-1}\nwmargintag{{\nwtagstyle{}\subpageref{NWwxjkf-Dyu9X-1}}}\moddef{topological sorting~{\nwtagstyle{}\subpageref{NWwxjkf-Dyu9X-1}}}\endmoddef\nwstartdeflinemarkup\nwusesondefline{\\{NWwxjkf-qfj8R-1}}\nwenddeflinemarkup
static void topsort_visit(SelfType* node,
                          std::vector<SelfType*>& order)\{

  if(node->topsort_status() != VISITED)\{

    \LA{}reset temporary parameters~{\nwtagstyle{}\subpageref{NWwxjkf-srI8w-1}}\RA{}
    \LA{}topsort visit children~{\nwtagstyle{}\subpageref{NWwxjkf-4SoTCX-1}}\RA{}
    \LA{}check exactness and add to order~{\nwtagstyle{}\subpageref{NWwxjkf-8ug4W-1}}\RA{}

    node->topsort_status() = VISITED;
  \} else \{
    node->has_multiple_references = true;
  \}
\}
\nwused{\\{NWwxjkf-qfj8R-1}}\nwendcode{}\nwbegindocs{114}%

The variable \impl{topsort\_status} is used to determine whether the node was already visited. In addition, if it was visited before, we now set \impl{has\_multiple\_references} to \impl{true}, which is used in computing the algebraic degree bound. Also, at the first visit, we reset all temporary parameters we will need in either the algebraic degree computation, or for maintaining the parents.
\nwenddocs{}\nwbegincode{115}\sublabel{NWwxjkf-srI8w-1}\nwmargintag{{\nwtagstyle{}\subpageref{NWwxjkf-srI8w-1}}}\moddef{reset temporary parameters~{\nwtagstyle{}\subpageref{NWwxjkf-srI8w-1}}}\endmoddef\nwstartdeflinemarkup\nwusesondefline{\\{NWwxjkf-Dyu9X-1}}\nwenddeflinemarkup
node->reset_algebraic_degree_parameters();
node->parents.clear();
node->dependency_count = 0;
\nwused{\\{NWwxjkf-Dyu9X-1}}\nwendcode{}\nwbegindocs{116}%

When the node has any children, we will also register it as a parent for them, if they are not exact. Otherwise they do not need to be recomputed and therefore will not be added to the list. So an inner node can have no dependencies, if all of its children are exact.
\nwenddocs{}\nwbegincode{117}\sublabel{NWwxjkf-4SoTCX-1}\nwmargintag{{\nwtagstyle{}\subpageref{NWwxjkf-4SoTCX-1}}}\moddef{topsort visit children~{\nwtagstyle{}\subpageref{NWwxjkf-4SoTCX-1}}}\endmoddef\nwstartdeflinemarkup\nwusesondefline{\\{NWwxjkf-Dyu9X-1}}\nwenddeflinemarkup
if(node->X) \{
  topsort_visit(node->X,order);
  if (!node->X->exact()) node->X->register_parent(node);
\}
if(node->Y) \{
  topsort_visit(node->Y,order);
  if (!node->Y->exact()) node->Y->register_parent(node);
\}
\nwused{\\{NWwxjkf-Dyu9X-1}}\nwendcode{}\nwbegindocs{118}%
Note that it is important to first visit the children and then call \impl{register\_parent}, since \impl{parents} will be reset at the first call of \impl{topsort\_visit}. Finally, all nodes which are not exact yet are added to \impl{order} for recomputation. If the node is exact, it does not need any recomputation and therefore will be treated as read-only during the evaluation process, so we need to initialize the separation bound right away.

\nwenddocs{}\nwbegincode{119}\sublabel{NWwxjkf-8ug4W-1}\nwmargintag{{\nwtagstyle{}\subpageref{NWwxjkf-8ug4W-1}}}\moddef{check exactness and add to order~{\nwtagstyle{}\subpageref{NWwxjkf-8ug4W-1}}}\endmoddef\nwstartdeflinemarkup\nwusesondefline{\\{NWwxjkf-Dyu9X-1}}\nwenddeflinemarkup
if(node->exact()) \{
  node->init_sep_bd();
\} else \{
  order.push_back(node);
\}
\nwused{\\{NWwxjkf-Dyu9X-1}}\nwendcode{}\nwbegindocs{120}%

\subsection{Class overview}

It follows the class definition. In contrast to \impl{sdag\_node} we do not use the custom memory management of \RealAlgebraic, since this is not guaranteed to be multithread-safe\footnote{The same is true for the bigfloat data, whose definition can be found in the appendix.}. We require \impl{boost} for using this class because the \impl{Thread\_manager} class relies on \impl{boost}, although it is not used directly in \impl{mtdag\_node}. So with an alternate thread manager, we could get rid of this requirement.
\nwenddocs{}\nwbegincode{121}\sublabel{NWwxjkf-qfj8R-1}\nwmargintag{{\nwtagstyle{}\subpageref{NWwxjkf-qfj8R-1}}}\moddef{mtdag\_node.hpp~{\nwtagstyle{}\subpageref{NWwxjkf-qfj8R-1}}}\endmoddef\nwstartdeflinemarkup\nwenddeflinemarkup
#ifndef RA_MTDAG_NODE_HPP
#define RA_MTDAG_NODE_HPP

#if (defined RA_CXX11 && defined RA_WITH_BOOST)

\LA{}includes~{\nwtagstyle{}\subpageref{NWwxjkf-EOZR2-1}}\RA{}

template <class Policies>
class mtdag_node\{ 
public:

  typedef mtdag_node<Policies> SelfType;

  \LA{}typedefs and enums~{\nwtagstyle{}\subpageref{NWwxjkf-1oEtvZ-1}}\RA{}

  \LA{}node data declaration~{\nwtagstyle{}\subpageref{NWwxjkf-feDHo-1}}\RA{}
  \LA{}node data getter and setter~{\nwtagstyle{}\subpageref{NWwxjkf-1qdLgq-1}}\RA{}
  
  \LA{}initialize bigfloat data~{\nwtagstyle{}\subpageref{NWwxjkf-4GuWSD-1}}\RA{}
  \LA{}topological sorting~{\nwtagstyle{}\subpageref{NWwxjkf-Dyu9X-1}}\RA{}
  
  \LA{}members~{\nwtagstyle{}\subpageref{NWwxjkf-3GwZ9F-1}}\RA{}
  \LA{}parent handling~{\nwtagstyle{}\subpageref{NWwxjkf-XH50e-1}}\RA{}
  \LA{}reference handling~{\nwtagstyle{}\subpageref{NWwxjkf-2obXTf-1}}\RA{}
  \LA{}node type~{\nwtagstyle{}\subpageref{NWwxjkf-2FFzRb-1}}\RA{}
  \LA{}threading members~{\nwtagstyle{}\subpageref{NWwxjkf-3eVWfJ-1}}\RA{}
  \LA{}algebraic degree computation members~{\nwtagstyle{}\subpageref{NWwxjkf-3R0KsL-1}}\RA{}
   
  \LA{}constructors~{\nwtagstyle{}\subpageref{NWwxjkf-56Gik-1}}\RA{}
  \LA{}destructor~{\nwtagstyle{}\subpageref{NWwxjkf-3BLOo7-1}}\RA{}
  
  \LA{}adjust hardware fp interval~{\nwtagstyle{}\subpageref{NWwxjkf-7WwWq-1}}\RA{}
  \LA{}fixed precision computation~{\nwtagstyle{}\subpageref{NWwxjkf-1xhPU9-1}}\RA{}
  
  \LA{}sign computation~{\nwtagstyle{}\subpageref{NWwxjkf-VFFA-1}}\RA{}

  \LA{}guaranteeing error~{\nwtagstyle{}\subpageref{NWwxjkf-xiB2Q-1}}\RA{}
  
  \LA{}separation bound computation~{\nwtagstyle{}\subpageref{NWwxjkf-3rifsU-1}}\RA{}
  \LA{}accuracy driven computation~{\nwtagstyle{}\subpageref{NWwxjkf-hyGrK-1}}\RA{}

  \LA{}bigfloat conversion~{\nwtagstyle{}\subpageref{NWwxjkf-16kevT-1}}\RA{}
\};

\LA{}static member definition~{\nwtagstyle{}\subpageref{NWwxjkf-2yBJ4v-1}}\RA{}

#endif//RA_CXX11 && RA_WITH_BOOST
#endif//RA_MTDAG_NODE_HPP
\nwnotused{mtdag\_node.hpp}\nwendcode{}\nwbegindocs{122}%

\bibliography{../../lit}
\bibliographystyle{abbrv}

\newpage
\appendix
\section{Appendix: Remaining code segments}

\paragraph{Class body}
\nwenddocs{}\nwbegincode{123}\sublabel{NWwxjkf-EOZR2-1}\nwmargintag{{\nwtagstyle{}\subpageref{NWwxjkf-EOZR2-1}}}\moddef{includes~{\nwtagstyle{}\subpageref{NWwxjkf-EOZR2-1}}}\endmoddef\nwstartdeflinemarkup\nwusesondefline{\\{NWwxjkf-qfj8R-1}}\nwenddeflinemarkup
#include "floating_point_basics.hpp"
#include "memory_allocation.hpp"
#include "error_handling.hpp"
#include "bits_and_pieces.hpp"
#include <cstdlib>
#include <vector>

#include "Thread_manager.hpp"
#include <mutex>
#include <condition_variable>
\nwused{\\{NWwxjkf-qfj8R-1}}\nwendcode{}\nwbegindocs{124}%

\nwenddocs{}\nwbegincode{125}\sublabel{NWwxjkf-1oEtvZ-1}\nwmargintag{{\nwtagstyle{}\subpageref{NWwxjkf-1oEtvZ-1}}}\moddef{typedefs and enums~{\nwtagstyle{}\subpageref{NWwxjkf-1oEtvZ-1}}}\endmoddef\nwstartdeflinemarkup\nwusesondefline{\\{NWwxjkf-qfj8R-1}}\nwenddeflinemarkup
typedef typename Policies::FilterPolicy         FilterPolicy;
typedef typename Policies::ApproximationPolicy  ApproximationPolicy;
typedef typename Policies::SeparationBound      SeparationBound;
typedef ApproximationPolicy        AP;
typedef typename AP::Approximation Approximation;
typedef typename AP::Exponent      Exponent;
typedef typename AP::Precision     Precision;

enum statustype\{CLEARED, VISITED\};

enum constructortype\{DOUBLE, BIGFLOAT,
                     NEGATION, ROOT, 
                     ADDITION, SUBTRACTION, MULTIPLICATION, DIVISION\};
\nwused{\\{NWwxjkf-qfj8R-1}}\nwendcode{}\nwbegindocs{126}%

\nwenddocs{}\nwbegincode{127}\sublabel{NWwxjkf-3GwZ9F-1}\nwmargintag{{\nwtagstyle{}\subpageref{NWwxjkf-3GwZ9F-1}}}\moddef{members~{\nwtagstyle{}\subpageref{NWwxjkf-3GwZ9F-1}}}\endmoddef\nwstartdeflinemarkup\nwusesondefline{\\{NWwxjkf-qfj8R-1}}\nwenddeflinemarkup
FilterPolicy I; 

Node_data *node_data;
SelfType *X, *Y;
\nwused{\\{NWwxjkf-qfj8R-1}}\nwendcode{}\nwbegindocs{128}%

\nwenddocs{}\nwbegincode{129}\sublabel{NWwxjkf-2FFzRb-1}\nwmargintag{{\nwtagstyle{}\subpageref{NWwxjkf-2FFzRb-1}}}\moddef{node type~{\nwtagstyle{}\subpageref{NWwxjkf-2FFzRb-1}}}\endmoddef\nwstartdeflinemarkup\nwusesondefline{\\{NWwxjkf-qfj8R-1}}\nwenddeflinemarkup
unsigned int node_type;

static inline unsigned int 
make_nt(unsigned int degree, unsigned int type)\{
  return (degree << 4) | type;
\}

inline int degree() const\{ return (node_type >> 4);\}
inline int contype() const\{ return (node_type & 15);\}
\nwused{\\{NWwxjkf-qfj8R-1}}\nwendcode{}\nwbegindocs{130}%

\paragraph{Bigfloat data}
\nwenddocs{}\nwbegincode{131}\sublabel{NWwxjkf-feDHo-1}\nwmargintag{{\nwtagstyle{}\subpageref{NWwxjkf-feDHo-1}}}\moddef{node data declaration~{\nwtagstyle{}\subpageref{NWwxjkf-feDHo-1}}}\endmoddef\nwstartdeflinemarkup\nwusesondefline{\\{NWwxjkf-qfj8R-1}}\nwenddeflinemarkup
struct Node_data\{ 

  \LA{}node data members~{\nwtagstyle{}\subpageref{NWwxjkf-Ai2gJ-1}}\RA{}   
  \LA{}node data constructors~{\nwtagstyle{}\subpageref{NWwxjkf-1sQxCF-1}}\RA{}

private:

  \LA{}ceil log2 double~{\nwtagstyle{}\subpageref{NWwxjkf-4V7g9k-1}}\RA{}
\};
\nwused{\\{NWwxjkf-qfj8R-1}}\nwendcode{}\nwbegindocs{132}%

\nwenddocs{}\nwbegincode{133}\sublabel{NWwxjkf-Ai2gJ-1}\nwmargintag{{\nwtagstyle{}\subpageref{NWwxjkf-Ai2gJ-1}}}\moddef{node data members~{\nwtagstyle{}\subpageref{NWwxjkf-Ai2gJ-1}}}\endmoddef\nwstartdeflinemarkup\nwusesondefline{\\{NWwxjkf-feDHo-1}}\nwenddeflinemarkup
Approximation approx;
Approximation error;

SeparationBound sep_bd;

Exponent requested_error;
statustype topsort_status;
\nwused{\\{NWwxjkf-feDHo-1}}\nwendcode{}\nwbegindocs{134}%

\nwenddocs{}\nwbegincode{135}\sublabel{NWwxjkf-1sQxCF-1}\nwmargintag{{\nwtagstyle{}\subpageref{NWwxjkf-1sQxCF-1}}}\moddef{node data constructors~{\nwtagstyle{}\subpageref{NWwxjkf-1sQxCF-1}}}\endmoddef\nwstartdeflinemarkup\nwusesondefline{\\{NWwxjkf-feDHo-1}}\nwenddeflinemarkup
Node_data():topsort_status(CLEARED)\{\}
  
Node_data(const double x,const double e)
       :approx(x),error(e),requested_error(e==0?0:ceil_log2(e)),
        topsort_status(CLEARED)\{\}
         
Node_data(const Approximation& x, 
          const double e)
       :approx(x),error(e),requested_error(e==0?0:ceil_log2(e)),
        topsort_status(CLEARED)\{\}
         
Node_data(const Approximation& x, 
          const Approximation& e)
       :approx(x),error(e),
        requested_error(AP::sign(error)==0?0:AP::ceil_log2(error)),
        topsort_status(CLEARED)\{\}
\nwused{\\{NWwxjkf-feDHo-1}}\nwendcode{}\nwbegindocs{136}%

\nwenddocs{}\nwbegincode{137}\sublabel{NWwxjkf-1qdLgq-1}\nwmargintag{{\nwtagstyle{}\subpageref{NWwxjkf-1qdLgq-1}}}\moddef{node data getter and setter~{\nwtagstyle{}\subpageref{NWwxjkf-1qdLgq-1}}}\endmoddef\nwstartdeflinemarkup\nwusesondefline{\\{NWwxjkf-qfj8R-1}}\nwenddeflinemarkup
\LA{}bigfloat getter and setter~{\nwtagstyle{}\subpageref{NWwxjkf-3Xx0J7-1}}\RA{}
\LA{}separation bound getter and setter~{\nwtagstyle{}\subpageref{NWwxjkf-2Ingyi-1}}\RA{}
\LA{}topological evaluation getter and setter~{\nwtagstyle{}\subpageref{NWwxjkf-bCnDP-1}}\RA{}
\nwused{\\{NWwxjkf-qfj8R-1}}\nwendcode{}\nwbegindocs{138}%

\nwenddocs{}\nwbegincode{139}\sublabel{NWwxjkf-3Xx0J7-1}\nwmargintag{{\nwtagstyle{}\subpageref{NWwxjkf-3Xx0J7-1}}}\moddef{bigfloat getter and setter~{\nwtagstyle{}\subpageref{NWwxjkf-3Xx0J7-1}}}\endmoddef\nwstartdeflinemarkup\nwusesondefline{\\{NWwxjkf-1qdLgq-1}}\nwenddeflinemarkup
inline Approximation& approx()\{  return node_data->approx;  \}
inline Approximation const& get_approx() const \{  return node_data->approx;  \}
inline Exponent ceil_log2_approx() const \{ return AP::ceil_log2(get_approx()); \}
inline Exponent floor_log2_approx() const \{ return AP::floor_log2(get_approx()); \}

inline Approximation& error()\{  return node_data->error;  \}
inline Approximation const& get_error() const \{ return node_data->error; \}
inline Exponent ceil_log2_error() const \{ return AP::ceil_log2(get_error()); \}

inline bool exact() const \{ return !static_cast<bool>(AP::sign(get_error())); \}
inline int sign() const \{ return AP::sign(get_approx()); \}
inline bool is_zero() const \{ return sign() == 0; \}
\nwused{\\{NWwxjkf-1qdLgq-1}}\nwendcode{}\nwbegindocs{140}%

\nwenddocs{}\nwbegincode{141}\sublabel{NWwxjkf-2Ingyi-2}\nwmargintag{{\nwtagstyle{}\subpageref{NWwxjkf-2Ingyi-2}}}\moddef{separation bound getter and setter~{\nwtagstyle{}\subpageref{NWwxjkf-2Ingyi-1}}}\plusendmoddef\nwstartdeflinemarkup\nwusesondefline{\\{NWwxjkf-1qdLgq-1}}\nwprevnextdefs{NWwxjkf-2Ingyi-1}{\relax}\nwenddeflinemarkup
inline SeparationBound& sep_bd()\{ return node_data->sep_bd; \}
\nwused{\\{NWwxjkf-1qdLgq-1}}\nwendcode{}\nwbegindocs{142}%

\nwenddocs{}\nwbegincode{143}\sublabel{NWwxjkf-bCnDP-1}\nwmargintag{{\nwtagstyle{}\subpageref{NWwxjkf-bCnDP-1}}}\moddef{topological evaluation getter and setter~{\nwtagstyle{}\subpageref{NWwxjkf-bCnDP-1}}}\endmoddef\nwstartdeflinemarkup\nwusesondefline{\\{NWwxjkf-1qdLgq-1}}\nwenddeflinemarkup
inline Exponent& requested_error()\{ return node_data->requested_error; \}
inline Exponent const& get_requested_error() const \{ return node_data->requested_error; \}

inline statustype& topsort_status()\{ return node_data->topsort_status; \}
\nwused{\\{NWwxjkf-1qdLgq-1}}\nwendcode{}\nwbegindocs{144}%

\nwenddocs{}\nwbegincode{145}\sublabel{NWwxjkf-4V7g9k-1}\nwmargintag{{\nwtagstyle{}\subpageref{NWwxjkf-4V7g9k-1}}}\moddef{ceil log2 double~{\nwtagstyle{}\subpageref{NWwxjkf-4V7g9k-1}}}\endmoddef\nwstartdeflinemarkup\nwusesondefline{\\{NWwxjkf-feDHo-1}}\nwenddeflinemarkup
static int ceil_log2(const double d) \{
  int k;
  std::frexp(d,&k);
  return k;
\}
\nwused{\\{NWwxjkf-feDHo-1}}\nwendcode{}\nwbegindocs{146}%

\nwenddocs{}\nwbegincode{147}\sublabel{NWwxjkf-4GuWSD-1}\nwmargintag{{\nwtagstyle{}\subpageref{NWwxjkf-4GuWSD-1}}}\moddef{initialize bigfloat data~{\nwtagstyle{}\subpageref{NWwxjkf-4GuWSD-1}}}\endmoddef\nwstartdeflinemarkup\nwusesondefline{\\{NWwxjkf-qfj8R-1}}\nwenddeflinemarkup
void init_node_data() \{  
  if (node_data) return;
  
  if (X) X->init_node_data();
  if (Y) Y->init_node_data();

  if(contype() == DOUBLE)\{
    node_data = new Node_data(I.get_point(),0.0);
  \} else if (I.is_finite())\{ 
    node_data = new Node_data(I.get_median(),I.get_radius());
    if (exact()) convert_to_bigfloat();
  \} else \{
    node_data = new Node_data();
    compute_op(AP::initial_prec());
  \}
\}
\nwused{\\{NWwxjkf-qfj8R-1}}\nwendcode{}\nwbegindocs{148}%

\paragraph{Convenience methods}

\nwenddocs{}\nwbegincode{149}\sublabel{NWwxjkf-16kevT-1}\nwmargintag{{\nwtagstyle{}\subpageref{NWwxjkf-16kevT-1}}}\moddef{bigfloat conversion~{\nwtagstyle{}\subpageref{NWwxjkf-16kevT-1}}}\endmoddef\nwstartdeflinemarkup\nwusesondefline{\\{NWwxjkf-qfj8R-1}}\nwenddeflinemarkup
inline void zeroize() \{
  I = FilterPolicy(0.0);
  AP::zeroize(approx());
  AP::zeroize(error());
\}

inline void convert_to_bigfloat() \{
  if (X && (X->ref_minus() == 0)) delete X;
  if (Y && (Y->ref_minus() == 0)) delete Y;
  X = Y = 0;
  node_type = make_nt(1,BIGFLOAT);
\}
\nwused{\\{NWwxjkf-qfj8R-1}}\nwendcode{}\nwbegindocs{150}%

\nwenddocs{}\nwbegincode{151}\sublabel{NWwxjkf-7WwWq-1}\nwmargintag{{\nwtagstyle{}\subpageref{NWwxjkf-7WwWq-1}}}\moddef{adjust hardware fp interval~{\nwtagstyle{}\subpageref{NWwxjkf-7WwWq-1}}}\endmoddef\nwstartdeflinemarkup\nwusesondefline{\\{NWwxjkf-qfj8R-1}}\nwenddeflinemarkup
void adjust_interval()\{
  AP::to_interval(I,get_approx());
  
  if(!exact())\{
    const double rad = AP::to_double(get_error(),AP::round_up());
    FilterPolicy err; 
    err.set_median_and_radius(0.0,rad);
    I += err;
  \}
\}
\nwused{\\{NWwxjkf-qfj8R-1}}\nwendcode{}\nwbegindocs{152}%

\nwenddocs{}\nwbegincode{153}\sublabel{NWwxjkf-35Jlkw-1}\nwmargintag{{\nwtagstyle{}\subpageref{NWwxjkf-35Jlkw-1}}}\moddef{adc auxillary functions~{\nwtagstyle{}\subpageref{NWwxjkf-35Jlkw-1}}}\endmoddef\nwstartdeflinemarkup\nwusesondefline{\\{NWwxjkf-hyGrK-1}}\nwenddeflinemarkup
inline Exponent ceil_log2_high() const \{
  if (is_zero()) return ceil_log2_error();
  if (exact()) return ceil_log2_approx();

  return static_cast<Exponent>(1) + std::max(ceil_log2_approx(),ceil_log2_error());
\}

inline Exponent floor_log2_low() const \{
  assert(!is_zero());
  if(exact()) return floor_log2_approx();

  assert(ceil_log2_error() < floor_log2_approx());
  return floor_log2_approx()-static_cast<Exponent>(1);
\}
\nwused{\\{NWwxjkf-hyGrK-1}}\nwendcode{}\nwbegindocs{154}%

\paragraph{Constructors and destructors}

\nwenddocs{}\nwbegincode{155}\sublabel{NWwxjkf-56Gik-1}\nwmargintag{{\nwtagstyle{}\subpageref{NWwxjkf-56Gik-1}}}\moddef{constructors~{\nwtagstyle{}\subpageref{NWwxjkf-56Gik-1}}}\endmoddef\nwstartdeflinemarkup\nwusesondefline{\\{NWwxjkf-qfj8R-1}}\nwenddeflinemarkup
mtdag_node(const FilterPolicy& i,Int_to_type<BIGFLOAT>)
:I(i),node_data(new Node_data()),X(0),Y(0),count(1),node_type(make_nt(1,BIGFLOAT))\{\}

mtdag_node(const double x)
:I(x),node_data(0),X(0),Y(0),count(1),node_type(make_nt(1,DOUBLE))\{
  assert(ra_isfinite(x));
\}

mtdag_node(const Approximation& x)
:node_data(new Node_data(x,0.0)),X(0),Y(0),count(1),node_type(make_nt(1,BIGFLOAT)) \{
  adjust_interval();
\}

mtdag_node(mtdag_node* a,mtdag_node* b,Int_to_type<ADDITION>)
:I( a->I + b->I ),node_data(0),X(a),Y(b),count(1),node_type(make_nt(1,ADDITION))\{
  \LA{}increase operands reference counter~{\nwtagstyle{}\subpageref{NWwxjkf-4NmrBg-1}}\RA{}
\}

mtdag_node(mtdag_node* a,mtdag_node* b,Int_to_type<SUBTRACTION>)
:I( a->I - b->I ),node_data(0),X(a),Y(b),count(1),node_type(make_nt(1,SUBTRACTION))\{
  \LA{}increase operands reference counter~{\nwtagstyle{}\subpageref{NWwxjkf-4NmrBg-1}}\RA{}
\}

mtdag_node(mtdag_node* a,mtdag_node* b,Int_to_type<MULTIPLICATION>)
:I( a->I * b->I ),node_data(0),X(a),Y(b),count(1),node_type(make_nt(1,MULTIPLICATION))\{
  \LA{}increase operands reference counter~{\nwtagstyle{}\subpageref{NWwxjkf-4NmrBg-1}}\RA{}
\}

mtdag_node(mtdag_node* a,mtdag_node* b,Int_to_type<DIVISION>)
:I( a->I / b->I ),node_data(0),X(a),Y(b),count(1),node_type(make_nt(1,DIVISION))\{
  \LA{}increase operands reference counter~{\nwtagstyle{}\subpageref{NWwxjkf-4NmrBg-1}}\RA{}
\}

mtdag_node(mtdag_node* a,const constructortype con)
:I(-(a->I)),node_data(0),X(a),Y(0),count(1),node_type(make_nt(1,con))\{ 
  assert(contype() == NEGATION);
  \LA{}increase single operand reference counter~{\nwtagstyle{}\subpageref{NWwxjkf-3ioqow-1}}\RA{}
\}

mtdag_node(mtdag_node* a,const int dd,const constructortype con)
:I(root(a->I,dd)),node_data(0),X(a),Y(0),count(1),node_type(make_nt(dd,con))\{
  assert(contype() == ROOT);
  assert(degree() == dd);
  assert(degree() >= 2);

  \LA{}increase single operand reference counter~{\nwtagstyle{}\subpageref{NWwxjkf-3ioqow-1}}\RA{}
\}
\nwused{\\{NWwxjkf-qfj8R-1}}\nwendcode{}\nwbegindocs{156}%

\nwenddocs{}\nwbegincode{157}\sublabel{NWwxjkf-4NmrBg-1}\nwmargintag{{\nwtagstyle{}\subpageref{NWwxjkf-4NmrBg-1}}}\moddef{increase operands reference counter~{\nwtagstyle{}\subpageref{NWwxjkf-4NmrBg-1}}}\endmoddef\nwstartdeflinemarkup\nwusesondefline{\\{NWwxjkf-56Gik-1}}\nwenddeflinemarkup
assert(X!=0); X->ref_plus();
assert(Y!=0); Y->ref_plus();
\nwused{\\{NWwxjkf-56Gik-1}}\nwendcode{}\nwbegindocs{158}%

\nwenddocs{}\nwbegincode{159}\sublabel{NWwxjkf-3ioqow-1}\nwmargintag{{\nwtagstyle{}\subpageref{NWwxjkf-3ioqow-1}}}\moddef{increase single operand reference counter~{\nwtagstyle{}\subpageref{NWwxjkf-3ioqow-1}}}\endmoddef\nwstartdeflinemarkup\nwusesondefline{\\{NWwxjkf-56Gik-1}}\nwenddeflinemarkup
assert(X!=0); X->ref_plus();
\nwused{\\{NWwxjkf-56Gik-1}}\nwendcode{}\nwbegindocs{160}%

\nwenddocs{}\nwbegincode{161}\sublabel{NWwxjkf-3BLOo7-1}\nwmargintag{{\nwtagstyle{}\subpageref{NWwxjkf-3BLOo7-1}}}\moddef{destructor~{\nwtagstyle{}\subpageref{NWwxjkf-3BLOo7-1}}}\endmoddef\nwstartdeflinemarkup\nwusesondefline{\\{NWwxjkf-qfj8R-1}}\nwenddeflinemarkup
~mtdag_node()\{
  delete node_data;

  if (X && (X->ref_minus() == 0)) delete X;
  if (Y && (Y->ref_minus() == 0)) delete Y;
\}
\nwused{\\{NWwxjkf-qfj8R-1}}\nwendcode{}\nwbegindocs{162}%

\paragraph{API methods}

\nwenddocs{}\nwbegincode{163}\sublabel{NWwxjkf-xiB2Q-1}\nwmargintag{{\nwtagstyle{}\subpageref{NWwxjkf-xiB2Q-1}}}\moddef{guaranteeing error~{\nwtagstyle{}\subpageref{NWwxjkf-xiB2Q-1}}}\endmoddef\nwstartdeflinemarkup\nwusesondefline{\\{NWwxjkf-qfj8R-1}}\nwenddeflinemarkup
\LA{}guaranteeing relative and absolute error~{\nwtagstyle{}\subpageref{NWwxjkf-Zc3P0-1}}\RA{}
\LA{}accessing internal approximation and error~{\nwtagstyle{}\subpageref{NWwxjkf-2wzHr2-1}}\RA{}
\nwused{\\{NWwxjkf-qfj8R-1}}\nwendcode{}\nwbegindocs{164}%

\nwenddocs{}\nwbegincode{165}\sublabel{NWwxjkf-Zc3P0-1}\nwmargintag{{\nwtagstyle{}\subpageref{NWwxjkf-Zc3P0-1}}}\moddef{guaranteeing relative and absolute error~{\nwtagstyle{}\subpageref{NWwxjkf-Zc3P0-1}}}\endmoddef\nwstartdeflinemarkup\nwusesondefline{\\{NWwxjkf-xiB2Q-1}}\nwprevnextdefs{\relax}{NWwxjkf-Zc3P0-2}\nwenddeflinemarkup
void guarantee_absolute_error_two_to(const Exponent& p)\{ 
  init_node_data();
  guarantee_bound_two_to(p); 
\}
\nwalsodefined{\\{NWwxjkf-Zc3P0-2}}\nwused{\\{NWwxjkf-xiB2Q-1}}\nwendcode{}\nwbegindocs{166}%

\nwenddocs{}\nwbegincode{167}\sublabel{NWwxjkf-Zc3P0-2}\nwmargintag{{\nwtagstyle{}\subpageref{NWwxjkf-Zc3P0-2}}}\moddef{guaranteeing relative and absolute error~{\nwtagstyle{}\subpageref{NWwxjkf-Zc3P0-1}}}\plusendmoddef\nwstartdeflinemarkup\nwusesondefline{\\{NWwxjkf-xiB2Q-1}}\nwprevnextdefs{NWwxjkf-Zc3P0-1}{\relax}\nwenddeflinemarkup
void guarantee_relative_error_two_to(const Exponent& p)\{ 
  if(sign_with_separation_bound()==0) return;
  init_node_data();
  if(exact()) return;
  
  Approximation tmp;
  AP::abs(tmp,get_approx(),AP::error_prec(),AP::round_down());
  AP::sub(tmp,tmp,get_error(),AP::error_prec(),AP::round_down());
  assert(AP::sign(tmp) > 0);
  
  const Exponent floor_log2_z_low(AP::floor_log2(tmp));
  const Exponent q(p + floor_log2_z_low);

  guarantee_bound_two_to(q);
\}
\nwused{\\{NWwxjkf-xiB2Q-1}}\nwendcode{}\nwbegindocs{168}%

\nwenddocs{}\nwbegincode{169}\sublabel{NWwxjkf-2wzHr2-1}\nwmargintag{{\nwtagstyle{}\subpageref{NWwxjkf-2wzHr2-1}}}\moddef{accessing internal approximation and error~{\nwtagstyle{}\subpageref{NWwxjkf-2wzHr2-1}}}\endmoddef\nwstartdeflinemarkup\nwusesondefline{\\{NWwxjkf-xiB2Q-1}}\nwprevnextdefs{\relax}{NWwxjkf-2wzHr2-2}\nwenddeflinemarkup
void get_approx_and_error(Approximation& a, Approximation& e)\{ 
  init_node_data();
  a = get_approx();
  e = get_error();
\}
\nwalsodefined{\\{NWwxjkf-2wzHr2-2}}\nwused{\\{NWwxjkf-xiB2Q-1}}\nwendcode{}\nwbegindocs{170}%

\nwenddocs{}\nwbegincode{171}\sublabel{NWwxjkf-2wzHr2-2}\nwmargintag{{\nwtagstyle{}\subpageref{NWwxjkf-2wzHr2-2}}}\moddef{accessing internal approximation and error~{\nwtagstyle{}\subpageref{NWwxjkf-2wzHr2-1}}}\plusendmoddef\nwstartdeflinemarkup\nwusesondefline{\\{NWwxjkf-xiB2Q-1}}\nwprevnextdefs{NWwxjkf-2wzHr2-1}{\relax}\nwenddeflinemarkup
const FilterPolicy& get_interval()\{
  if(contype()==DOUBLE) return I;
  init_node_data();
  adjust_interval();
  return I;
\}
\nwused{\\{NWwxjkf-xiB2Q-1}}\nwendcode{}\nwbegindocs{172}%

\paragraph{Separation bound computation}

\nwenddocs{}\nwbegincode{173}\sublabel{NWwxjkf-4Litu7-1}\nwmargintag{{\nwtagstyle{}\subpageref{NWwxjkf-4Litu7-1}}}\moddef{separation bound concrete computation~{\nwtagstyle{}\subpageref{NWwxjkf-4Litu7-1}}}\endmoddef\nwstartdeflinemarkup\nwusesondefline{\\{NWwxjkf-3rifsU-1}}\nwenddeflinemarkup
void init_sep_bd()\{

  switch(contype()) \{ 
    case DOUBLE:      
      assert(I.is_singleton());
      sep_bd().set(I.get_point());
      break;   
    case BIGFLOAT:
      sep_bd().set(get_approx());
      break;
    case NEGATION:
      sep_bd().negation(X->sep_bd());
      break;
    case ROOT:
      sep_bd().root(X->sep_bd(),degree());
      break;
    case ADDITION:
      sep_bd().addition(X->sep_bd(),Y->sep_bd());
      break;
    case SUBTRACTION:
      sep_bd().subtraction(X->sep_bd(),Y->sep_bd());
      break;
    case MULTIPLICATION:
      sep_bd().multiplication(X->sep_bd(),Y->sep_bd());
      break;
    case DIVISION:
      sep_bd().division(X->sep_bd(),Y->sep_bd());
      break;
  \}
\}
\nwused{\\{NWwxjkf-3rifsU-1}}\nwendcode{}\nwbegindocs{174}%

\nwenddocs{}\nwbegincode{175}\sublabel{NWwxjkf-VFFA-1}\nwmargintag{{\nwtagstyle{}\subpageref{NWwxjkf-VFFA-1}}}\moddef{sign computation~{\nwtagstyle{}\subpageref{NWwxjkf-VFFA-1}}}\endmoddef\nwstartdeflinemarkup\nwusesondefline{\\{NWwxjkf-qfj8R-1}}\nwenddeflinemarkup
\LA{}sign computation wrapper~{\nwtagstyle{}\subpageref{NWwxjkf-9mRa2-1}}\RA{}
\LA{}sign computation auxillary methods~{\nwtagstyle{}\subpageref{NWwxjkf-1M0F4U-1}}\RA{}
\nwused{\\{NWwxjkf-qfj8R-1}}\nwendcode{}\nwbegindocs{176}%

\nwenddocs{}\nwbegincode{177}\sublabel{NWwxjkf-9mRa2-1}\nwmargintag{{\nwtagstyle{}\subpageref{NWwxjkf-9mRa2-1}}}\moddef{sign computation wrapper~{\nwtagstyle{}\subpageref{NWwxjkf-9mRa2-1}}}\endmoddef\nwstartdeflinemarkup\nwusesondefline{\\{NWwxjkf-VFFA-1}}\nwenddeflinemarkup
int sign_with_separation_bound()\{
  if(!I.contains(0.0))\{
    return I.get_point() > 0.0 ? 1 : -1;
  \}else if(I.is_singleton())\{
    return 0;
  \}

  init_node_data(); 
  separate_from_zero();

  return sign();
\}
\nwused{\\{NWwxjkf-VFFA-1}}\nwendcode{}\nwbegindocs{178}%

\nwenddocs{}\nwbegincode{179}\sublabel{NWwxjkf-1M0F4U-1}\nwmargintag{{\nwtagstyle{}\subpageref{NWwxjkf-1M0F4U-1}}}\moddef{sign computation auxillary methods~{\nwtagstyle{}\subpageref{NWwxjkf-1M0F4U-1}}}\endmoddef\nwstartdeflinemarkup\nwusesondefline{\\{NWwxjkf-VFFA-1}}\nwenddeflinemarkup
inline void separate_from_zero() \{
  if (!exact() && (is_zero() || floor_log2_approx() <= ceil_log2_error())) \{
    Exponent log2_abserr = ceil_log2_error(); 
    Exponent log2_relerr = -27;                          

    do \{  
      log2_relerr *= 2;
      log2_abserr += log2_relerr;

      guarantee_bound_two_to(log2_abserr);
    \} while(!exact() && 
            (is_zero() || floor_log2_approx() <= ceil_log2_error()));
  \}
\}
\nwused{\\{NWwxjkf-VFFA-1}}\nwendcode{}\nwbegindocs{180}%

\paragraph{Fixed precision computation}

\nwenddocs{}\nwbegincode{181}\sublabel{NWwxjkf-1xhPU9-1}\nwmargintag{{\nwtagstyle{}\subpageref{NWwxjkf-1xhPU9-1}}}\moddef{fixed precision computation~{\nwtagstyle{}\subpageref{NWwxjkf-1xhPU9-1}}}\endmoddef\nwstartdeflinemarkup\nwusesondefline{\\{NWwxjkf-qfj8R-1}}\nwenddeflinemarkup
\LA{}fixed precision wrapper~{\nwtagstyle{}\subpageref{NWwxjkf-3EYhwl-1}}\RA{}
\LA{}fixed precision routines~{\nwtagstyle{}\subpageref{NWwxjkf-1F8iKo-1}}\RA{}
\nwused{\\{NWwxjkf-qfj8R-1}}\nwendcode{}\nwbegindocs{182}%

\nwenddocs{}\nwbegincode{183}\sublabel{NWwxjkf-3EYhwl-1}\nwmargintag{{\nwtagstyle{}\subpageref{NWwxjkf-3EYhwl-1}}}\moddef{fixed precision wrapper~{\nwtagstyle{}\subpageref{NWwxjkf-3EYhwl-1}}}\endmoddef\nwstartdeflinemarkup\nwusesondefline{\\{NWwxjkf-1xhPU9-1}}\nwenddeflinemarkup
void compute_op(const Precision p)\{ 

  switch(contype())\{
    case DOUBLE:
      assert(I.is_singleton());
      approx() = Approximation(I.get_point());
      assert(exact());
      return;
    case BIGFLOAT:
      assert(exact());
      return;
    case NEGATION:        compute_neg();    break;
    case ROOT:            compute_root(p); break;
    case ADDITION:        compute_add(p);  break;
    case SUBTRACTION:     compute_sub(p);  break;
    case MULTIPLICATION:  compute_mul(p);  break;
    case DIVISION:        compute_div(p);  break;
  \}
  
  if(exact())\{
    convert_to_bigfloat();
  \}
  else \{
    requested_error() = ceil_log2_error();
  \}

  \LA{}try to improve filter policy~{\nwtagstyle{}\subpageref{NWwxjkf-KEXze-1}}\RA{}
\}
\nwused{\\{NWwxjkf-1xhPU9-1}}\nwendcode{}\nwbegindocs{184}%

\nwenddocs{}\nwbegincode{185}\sublabel{NWwxjkf-KEXze-1}\nwmargintag{{\nwtagstyle{}\subpageref{NWwxjkf-KEXze-1}}}\moddef{try to improve filter policy~{\nwtagstyle{}\subpageref{NWwxjkf-KEXze-1}}}\endmoddef\nwstartdeflinemarkup\nwusesondefline{\\{NWwxjkf-3j4HBX-1}\\{NWwxjkf-3EYhwl-1}}\nwenddeflinemarkup
if(I.can_be_improved()) adjust_interval();
\nwused{\\{NWwxjkf-3j4HBX-1}\\{NWwxjkf-3EYhwl-1}}\nwendcode{}\nwbegindocs{186}%

\nwenddocs{}\nwbegincode{187}\sublabel{NWwxjkf-1F8iKo-1}\nwmargintag{{\nwtagstyle{}\subpageref{NWwxjkf-1F8iKo-1}}}\moddef{fixed precision routines~{\nwtagstyle{}\subpageref{NWwxjkf-1F8iKo-1}}}\endmoddef\nwstartdeflinemarkup\nwusesondefline{\\{NWwxjkf-1xhPU9-1}}\nwenddeflinemarkup
inline void compute_neg() \{
  AP::neg(approx(),X->get_approx(),AP::get_prec(X->get_approx()),AP::round_to_nearest());
  error() = X->get_error();
\}

void compute_add(const Precision p)\{
  assert(contype() == ADDITION);
  const bool isexact = 
  AP::add(approx(),X->get_approx(),Y->get_approx(),p,AP::round_to_nearest());

  AP::add(error(),X->get_error(),Y->get_error(),AP::error_prec(),AP::round_up());
  Approximation tmp;
  \LA{}compute and add operation error~{\nwtagstyle{}\subpageref{NWwxjkf-1o2REi-1}}\RA{}   
\}

void compute_sub(const Precision p)\{
  assert(contype() == SUBTRACTION);
  const bool isexact = 
  AP::sub(approx(),X->get_approx(),Y->get_approx(),p,AP::round_to_nearest());

  AP::add(error(),X->get_error(),Y->get_error(),AP::error_prec(),AP::round_up());
  Approximation tmp;
  \LA{}compute and add operation error~{\nwtagstyle{}\subpageref{NWwxjkf-1o2REi-1}}\RA{}
\}

void compute_mul(const Precision p)\{
  assert(contype() == MULTIPLICATION);
  const bool isexact =
  AP::mul(approx(),X->get_approx(),Y->get_approx(),p,AP::round_to_nearest());

  Approximation tmp;
  AP::abs(tmp,X->get_approx(),AP::error_prec(),AP::round_up());  
  AP::mul(error(),tmp,Y->get_error(),AP::error_prec(),AP::round_up());
     
  AP::abs(tmp,Y->get_approx(),AP::error_prec(),AP::round_up());
  AP::add(tmp,tmp,Y->get_error(),AP::error_prec(),AP::round_up());
  AP::mul(tmp,tmp,X->get_error(),AP::error_prec(),AP::round_up());
    
  AP::add(error(),get_error(),tmp,AP::error_prec(),AP::round_up());
  \LA{}compute and add operation error~{\nwtagstyle{}\subpageref{NWwxjkf-1o2REi-1}}\RA{}
\}

void compute_div(const Precision p)\{
  assert(contype() == DIVISION);
  
  if(Y->sign_with_separation_bound() == 0)
    real_algebraic_error_handler(1,__FILE__,__LINE__,RA_FUNCTION,
                                 "Division by zero.");
          
  const bool isexact = 
  AP::div(approx(),X->get_approx(),Y->get_approx(),p,AP::round_to_nearest());

  Approximation tmp;
  AP::abs(tmp,Y->get_approx(),AP::error_prec(),AP::round_down());
  AP::sub(tmp,tmp,Y->get_error(),AP::error_prec(),AP::round_down());
  assert(AP::sign(tmp) > 0);
  Exponent minus_floor_log2_ylow = -AP::floor_log2(tmp);
\}

void compute_root(const Precision p)\{
  assert(contype() == ROOT);
  
  if(X->sign_with_separation_bound() < 0)
    real_algebraic_error_handler(1,__FILE__,__LINE__,RA_FUNCTION,
                                 "Root of a negative number.");
    
  const bool isexact = 
  AP::root(approx(),X->get_approx(),degree(),p,AP::round_to_nearest());

  if(X->is_zero())\{
    assert(X->exact());
    AP::zeroize(error());
  \} else \{
    Approximation tmp;
    AP::sub(tmp,X->get_approx(),X->get_error(),AP::error_prec(),AP::round_down());
    assert(AP::sign(tmp) > 0);
    
    const int d = degree();
    int one_plus_floor_log2_d;
    std::frexp(static_cast<double>(d),&one_plus_floor_log2_d);

    Exponent log2_error_term = 
       (static_cast<Exponent>(1-d) * AP::floor_log2(tmp)) / static_cast<Exponent>(d) 
      + static_cast<Exponent>( 2 - one_plus_floor_log2_d) ;
    AP::ldexp(error(),X->get_error(),log2_error_term);
    \LA{}compute and add operation error~{\nwtagstyle{}\subpageref{NWwxjkf-1o2REi-1}}\RA{}
  \}
\}
\nwused{\\{NWwxjkf-1xhPU9-1}}\nwendcode{}\nwbegindocs{188}%

\nwenddocs{}\nwbegincode{189}\sublabel{NWwxjkf-1o2REi-1}\nwmargintag{{\nwtagstyle{}\subpageref{NWwxjkf-1o2REi-1}}}\moddef{compute and add operation error~{\nwtagstyle{}\subpageref{NWwxjkf-1o2REi-1}}}\endmoddef\nwstartdeflinemarkup\nwusesondefline{\\{NWwxjkf-1F8iKo-1}}\nwenddeflinemarkup
if(!isexact)\{
  AP::abs(tmp,get_approx(),AP::error_prec(),AP::round_up());
  AP::ldexp(tmp,tmp,-static_cast<Exponent>(p));
  AP::add(error(),get_error(),tmp,AP::error_prec(),AP::round_up());
\}
\nwused{\\{NWwxjkf-1F8iKo-1}}\nwendcode{}\nwbegindocs{190}%

\paragraph{Accuracy-driven computation overview}

\nwenddocs{}\nwbegincode{191}\sublabel{NWwxjkf-hyGrK-1}\nwmargintag{{\nwtagstyle{}\subpageref{NWwxjkf-hyGrK-1}}}\moddef{accuracy driven computation~{\nwtagstyle{}\subpageref{NWwxjkf-hyGrK-1}}}\endmoddef\nwstartdeflinemarkup\nwusesondefline{\\{NWwxjkf-qfj8R-1}}\nwenddeflinemarkup
\LA{}adc wrapper~{\nwtagstyle{}\subpageref{NWwxjkf-1RZNmO-1}}\RA{}
\LA{}adc error propagation~{\nwtagstyle{}\subpageref{NWwxjkf-4EGrT7-1}}\RA{}
\LA{}adc recomputation~{\nwtagstyle{}\subpageref{NWwxjkf-ZVr1v-1}}\RA{}
\LA{}adc fixup~{\nwtagstyle{}\subpageref{NWwxjkf-3j4HBX-1}}\RA{}
\LA{}adc auxillary functions~{\nwtagstyle{}\subpageref{NWwxjkf-35Jlkw-1}}\RA{}
\nwused{\\{NWwxjkf-qfj8R-1}}\nwendcode{}\nwbegindocs{192}%

\paragraph{Error propagation for accuracy-driven computation}

\nwenddocs{}\nwbegincode{193}\sublabel{NWwxjkf-6vwGL-1}\nwmargintag{{\nwtagstyle{}\subpageref{NWwxjkf-6vwGL-1}}}\moddef{propagate error topologically descending~{\nwtagstyle{}\subpageref{NWwxjkf-6vwGL-1}}}\endmoddef\nwstartdeflinemarkup\nwusesondefline{\\{NWwxjkf-1RZNmO-1}}\nwenddeflinemarkup
for(std::vector<SelfType*>::const_reverse_iterator it=order.rbegin();it!=order.rend();++it)
\{
  assert((*it)->contype() != DOUBLE);
  assert((*it)->contype() != BIGFLOAT);

  switch((*it)->contype()) \{ 
    case NEGATION:       (*it)->prop_neg_error();  break;
    case ROOT:           (*it)->prop_root_error(); break;
    case ADDITION:
    case SUBTRACTION:    (*it)->prop_add_or_sub_error();  break;
    case MULTIPLICATION: (*it)->prop_mul_error();  break;
    case DIVISION:       (*it)->prop_div_error();  break;
  \}    
\}   
\nwused{\\{NWwxjkf-1RZNmO-1}}\nwendcode{}\nwbegindocs{194}%

\nwenddocs{}\nwbegincode{195}\sublabel{NWwxjkf-1alRwG-1}\nwmargintag{{\nwtagstyle{}\subpageref{NWwxjkf-1alRwG-1}}}\moddef{set q and check exactness~{\nwtagstyle{}\subpageref{NWwxjkf-1alRwG-1}}}\endmoddef\nwstartdeflinemarkup\nwusesondefline{\\{NWwxjkf-4EGrT7-1}\\{NWwxjkf-ZVr1v-1}\\{NWwxjkf-ZVr1v-2}}\nwenddeflinemarkup
assert(!exact());

const Exponent q = get_requested_error();
if(ceil_log2_error() <= q) return;
\nwused{\\{NWwxjkf-4EGrT7-1}\\{NWwxjkf-ZVr1v-1}\\{NWwxjkf-ZVr1v-2}}\nwendcode{}\nwbegindocs{196}%

\nwenddocs{}\nwbegincode{197}\sublabel{NWwxjkf-4EGrT7-1}\nwmargintag{{\nwtagstyle{}\subpageref{NWwxjkf-4EGrT7-1}}}\moddef{adc error propagation~{\nwtagstyle{}\subpageref{NWwxjkf-4EGrT7-1}}}\endmoddef\nwstartdeflinemarkup\nwusesondefline{\\{NWwxjkf-hyGrK-1}}\nwenddeflinemarkup
void prop_neg_error()\{
  assert(contype() == NEGATION);
  
  \LA{}set q and check exactness~{\nwtagstyle{}\subpageref{NWwxjkf-1alRwG-1}}\RA{}
  
  if(q < X->get_requested_error()) X->requested_error() = q;
\}

void prop_root_error()\{
  assert(contype() == ROOT);
  
  \LA{}set q and check exactness~{\nwtagstyle{}\subpageref{NWwxjkf-1alRwG-1}}\RA{}
  
  if( !X->is_zero() )\{
    const Exponent floor_log2_xlow = X->floor_log2_low();

    const int d = degree();
    int one_plus_floor_log2_d;
    std::frexp(static_cast<double>(d),&one_plus_floor_log2_d);

    const Exponent qx = 
    std::min(floor_log2_xlow - static_cast<Exponent>(1),
            (static_cast<Exponent>(d-1) * floor_log2_xlow) / static_cast<Exponent>(d)
             + q + static_cast<Exponent>(one_plus_floor_log2_d - 3)); 
    
    if(qx < X->get_requested_error()) X->requested_error() = qx;
  \}
\}

void prop_add_or_sub_error()\{
  assert(contype() == ADDITION || contype() == SUBTRACTION);
  
  \LA{}set q and check exactness~{\nwtagstyle{}\subpageref{NWwxjkf-1alRwG-1}}\RA{}
  
  const Exponent qx = q - static_cast<Exponent>(2);
  if(qx < X->get_requested_error()) X->requested_error() = qx;
  if(qx < Y->get_requested_error()) Y->requested_error() = qx;
\}

void prop_mul_error()\{
  assert(contype() == MULTIPLICATION);
  
  \LA{}set q and check exactness~{\nwtagstyle{}\subpageref{NWwxjkf-1alRwG-1}}\RA{}
  
  if ((X->is_zero() && X->exact()) ||
      (Y->is_zero() && Y->exact())) return;

  const Exponent two(2);
  const Exponent c = q - two; 

  Exponent qx = c - Y->ceil_log2_high();
  Exponent qy = c - X->ceil_log2_high();

  if(qx + qy > c)\{
    qx = (c+qx-qy)/two;
    qy = c - qx;
  \}

  assert(qx <= q - static_cast<Exponent>(2) - Y->ceil_log2_high());
  assert(qy <= q - static_cast<Exponent>(2) - X->ceil_log2_high());
  assert(qx+qy <= q - static_cast<Exponent>(2));

  if(qx < X->get_requested_error()) X->requested_error() = qx;
  if(qy < Y->get_requested_error()) Y->requested_error() = qy;
\}

void prop_div_error()\{
  assert(contype() == DIVISION);
  
  \LA{}set q and check exactness~{\nwtagstyle{}\subpageref{NWwxjkf-1alRwG-1}}\RA{}
  
  if(X->is_zero() && X->exact()) return;
  
  const Exponent floor_log2_ylow = Y->floor_log2_low();

  const Exponent qx = q - static_cast<Exponent>(4) + floor_log2_ylow;
  const Exponent qy = std::min(floor_log2_ylow - static_cast<Exponent>(1),
                               q - static_cast<Exponent>(4) - X->ceil_log2_high() 
                                  + static_cast<Exponent>(2)*floor_log2_ylow);
  
  if(qx < X->get_requested_error()) X->requested_error() = qx;
  if(qy < Y->get_requested_error()) Y->requested_error() = qy;
\}
\nwused{\\{NWwxjkf-hyGrK-1}}\nwendcode{}\nwbegindocs{198}%

\paragraph{Recomputation methods for accuracy-driven computation}

\nwenddocs{}\nwbegincode{199}\sublabel{NWwxjkf-ZVr1v-1}\nwmargintag{{\nwtagstyle{}\subpageref{NWwxjkf-ZVr1v-1}}}\moddef{adc recomputation~{\nwtagstyle{}\subpageref{NWwxjkf-ZVr1v-1}}}\endmoddef\nwstartdeflinemarkup\nwusesondefline{\\{NWwxjkf-hyGrK-1}}\nwprevnextdefs{\relax}{NWwxjkf-ZVr1v-2}\nwenddeflinemarkup
void recompute_neg()\{
  assert(contype() == NEGATION);
  
  \LA{}set q and check exactness~{\nwtagstyle{}\subpageref{NWwxjkf-1alRwG-1}}\RA{}

  AP::neg(approx(),X->get_approx(),AP::get_prec(X->get_approx()),AP::round_to_nearest());
  error() = X->get_error();    
\}

void recompute_root()\{
  assert(contype() == ROOT);
  
  \LA{}set q and check exactness~{\nwtagstyle{}\subpageref{NWwxjkf-1alRwG-1}}\RA{}

  Precision p(AP::min_prec());
  int d = degree();
  
  if( !X->is_zero() )\{
    const Exponent relerr = X->ceil_log2_approx()/static_cast<Exponent>(d) 
                                                + static_cast<Exponent>(2) - q;
    p = AP::convert_to_prec(relerr);
  \}
  
  const bool isexact = AP::root(approx(),X->get_approx(),d,p,AP::round_to_nearest());

  \LA{}set new error for unary operators~{\nwtagstyle{}\subpageref{NWwxjkf-BA3wG-1}}\RA{}
\}

void recompute_add()\{
  assert(contype() == ADDITION);
  
  \LA{}set q and check exactness~{\nwtagstyle{}\subpageref{NWwxjkf-1alRwG-1}}\RA{} 
  \LA{}set p for addition and subtraction~{\nwtagstyle{}\subpageref{NWwxjkf-fTSXM-1}}\RA{}
  
  const bool isexact = 
  AP::add(approx(),X->get_approx(),Y->get_approx(),p,AP::round_to_nearest());    
  
  \LA{}set new error for binary operators~{\nwtagstyle{}\subpageref{NWwxjkf-2zgPbv-1}}\RA{}
\}

void recompute_sub()\{
  assert(contype() == SUBTRACTION);
  
  \LA{}set q and check exactness~{\nwtagstyle{}\subpageref{NWwxjkf-1alRwG-1}}\RA{}
  \LA{}set p for addition and subtraction~{\nwtagstyle{}\subpageref{NWwxjkf-fTSXM-1}}\RA{}
  
  const bool isexact = 
  AP::sub(approx(),X->get_approx(),Y->get_approx(),p,AP::round_to_nearest());    
  
  \LA{}set new error for binary operators~{\nwtagstyle{}\subpageref{NWwxjkf-2zgPbv-1}}\RA{}
\}
\nwalsodefined{\\{NWwxjkf-ZVr1v-2}}\nwused{\\{NWwxjkf-hyGrK-1}}\nwendcode{}\nwbegindocs{200}%

\nwenddocs{}\nwbegincode{201}\sublabel{NWwxjkf-fTSXM-1}\nwmargintag{{\nwtagstyle{}\subpageref{NWwxjkf-fTSXM-1}}}\moddef{set p for addition and subtraction~{\nwtagstyle{}\subpageref{NWwxjkf-fTSXM-1}}}\endmoddef\nwstartdeflinemarkup\nwusesondefline{\\{NWwxjkf-ZVr1v-1}}\nwenddeflinemarkup
Precision p(AP::min_prec());
if (!X->is_zero() && !Y->is_zero()) \{
  const Exponent relerr = std::max(X->ceil_log2_approx(),
                             Y->ceil_log2_approx())
                                                + static_cast<Exponent>(2) - q;
  p = AP::convert_to_prec(relerr); 
\} else if (!X->is_zero()) \{
  const Exponent relerr = X->ceil_log2_approx() + static_cast<Exponent>(1) - q;
  p = AP::convert_to_prec(relerr);
\} else if (!Y->is_zero()) \{
  const Exponent relerr = Y->ceil_log2_approx() + static_cast<Exponent>(1) - q;
  p = AP::convert_to_prec(relerr);
\}
\nwused{\\{NWwxjkf-ZVr1v-1}}\nwendcode{}\nwbegindocs{202}%

\nwenddocs{}\nwbegincode{203}\sublabel{NWwxjkf-ZVr1v-2}\nwmargintag{{\nwtagstyle{}\subpageref{NWwxjkf-ZVr1v-2}}}\moddef{adc recomputation~{\nwtagstyle{}\subpageref{NWwxjkf-ZVr1v-1}}}\plusendmoddef\nwstartdeflinemarkup\nwusesondefline{\\{NWwxjkf-hyGrK-1}}\nwprevnextdefs{NWwxjkf-ZVr1v-1}{\relax}\nwenddeflinemarkup
void recompute_mul()\{
  assert(contype() == MULTIPLICATION);
  
  \LA{}set q and check exactness~{\nwtagstyle{}\subpageref{NWwxjkf-1alRwG-1}}\RA{}
  
  bool isexact = false; 
  if( !X->is_zero() && !Y->is_zero() )\{
    const Exponent two(2);
    const Exponent relerr = X->ceil_log2_approx()
                      + Y->ceil_log2_approx() + two - q;
                    
    const Precision p = AP::convert_to_prec(relerr);
    
    isexact =
    AP::mul(approx(),X->get_approx(),Y->get_approx(),p,AP::round_to_nearest());
  \}
  else
  \{
    isexact = true;
    AP::zeroize(approx());
  \}    
  
  \LA{}set new error for binary operators~{\nwtagstyle{}\subpageref{NWwxjkf-2zgPbv-1}}\RA{}
\}

void recompute_div()\{
  assert(contype() == DIVISION);
  
  \LA{}set q and check exactness~{\nwtagstyle{}\subpageref{NWwxjkf-1alRwG-1}}\RA{}
  
  assert(!Y->is_zero());

  bool isexact = false; 
  if( !X->is_zero() )\{
    const Exponent relerr = X->ceil_log2_approx() 
                            - Y->floor_log2_approx() + static_cast<Exponent>(1) - q;
    const Precision p = AP::convert_to_prec(relerr);
    
    isexact = 
    AP::div(approx(),X->get_approx(),Y->get_approx(),p,AP::round_to_nearest());
  \}
  else
  \{
    AP::zeroize(approx());
    isexact = true;
  \}
  
  \LA{}set new error for binary operators~{\nwtagstyle{}\subpageref{NWwxjkf-2zgPbv-1}}\RA{}
\}
\nwused{\\{NWwxjkf-hyGrK-1}}\nwendcode{}\nwbegindocs{204}%

\nwenddocs{}\nwbegincode{205}\sublabel{NWwxjkf-BA3wG-1}\nwmargintag{{\nwtagstyle{}\subpageref{NWwxjkf-BA3wG-1}}}\moddef{set new error for unary operators~{\nwtagstyle{}\subpageref{NWwxjkf-BA3wG-1}}}\endmoddef\nwstartdeflinemarkup\nwusesondefline{\\{NWwxjkf-ZVr1v-1}}\nwenddeflinemarkup
if ( isexact && X->exact() ) AP::zeroize(error());   
else AP::pow2(error(),q);
\nwused{\\{NWwxjkf-ZVr1v-1}}\nwendcode{}\nwbegindocs{206}%

\nwenddocs{}\nwbegincode{207}\sublabel{NWwxjkf-2zgPbv-1}\nwmargintag{{\nwtagstyle{}\subpageref{NWwxjkf-2zgPbv-1}}}\moddef{set new error for binary operators~{\nwtagstyle{}\subpageref{NWwxjkf-2zgPbv-1}}}\endmoddef\nwstartdeflinemarkup\nwusesondefline{\\{NWwxjkf-ZVr1v-1}\\{NWwxjkf-ZVr1v-2}}\nwenddeflinemarkup
if ( isexact && X->exact() && Y->exact() ) AP::zeroize(error());   
else AP::pow2(error(),q);
\nwused{\\{NWwxjkf-ZVr1v-1}\\{NWwxjkf-ZVr1v-2}}\nwendcode{}

\nwixlogsorted{c}{{accessing internal approximation and error}{NWwxjkf-2wzHr2-1}{\nwixu{NWwxjkf-xiB2Q-1}\nwixd{NWwxjkf-2wzHr2-1}\nwixd{NWwxjkf-2wzHr2-2}}}%
\nwixlogsorted{c}{{accuracy driven computation}{NWwxjkf-hyGrK-1}{\nwixu{NWwxjkf-qfj8R-1}\nwixd{NWwxjkf-hyGrK-1}}}%
\nwixlogsorted{c}{{adc auxillary functions}{NWwxjkf-35Jlkw-1}{\nwixd{NWwxjkf-35Jlkw-1}\nwixu{NWwxjkf-hyGrK-1}}}%
\nwixlogsorted{c}{{adc error propagation}{NWwxjkf-4EGrT7-1}{\nwixu{NWwxjkf-hyGrK-1}\nwixd{NWwxjkf-4EGrT7-1}}}%
\nwixlogsorted{c}{{adc fixup}{NWwxjkf-3j4HBX-1}{\nwixd{NWwxjkf-3j4HBX-1}\nwixu{NWwxjkf-hyGrK-1}}}%
\nwixlogsorted{c}{{adc recomputation}{NWwxjkf-ZVr1v-1}{\nwixu{NWwxjkf-hyGrK-1}\nwixd{NWwxjkf-ZVr1v-1}\nwixd{NWwxjkf-ZVr1v-2}}}%
\nwixlogsorted{c}{{adc wrapper}{NWwxjkf-1RZNmO-1}{\nwixd{NWwxjkf-1RZNmO-1}\nwixd{NWwxjkf-1RZNmO-2}\nwixu{NWwxjkf-hyGrK-1}}}%
\nwixlogsorted{c}{{adjust hardware fp interval}{NWwxjkf-7WwWq-1}{\nwixu{NWwxjkf-qfj8R-1}\nwixd{NWwxjkf-7WwWq-1}}}%
\nwixlogsorted{c}{{algebraic degree computation members}{NWwxjkf-3R0KsL-1}{\nwixd{NWwxjkf-3R0KsL-1}\nwixd{NWwxjkf-3R0KsL-2}\nwixd{NWwxjkf-3R0KsL-3}\nwixu{NWwxjkf-qfj8R-1}}}%
\nwixlogsorted{c}{{bigfloat conversion}{NWwxjkf-16kevT-1}{\nwixu{NWwxjkf-qfj8R-1}\nwixd{NWwxjkf-16kevT-1}}}%
\nwixlogsorted{c}{{bigfloat getter and setter}{NWwxjkf-3Xx0J7-1}{\nwixu{NWwxjkf-1qdLgq-1}\nwixd{NWwxjkf-3Xx0J7-1}}}%
\nwixlogsorted{c}{{ceil log2 double}{NWwxjkf-4V7g9k-1}{\nwixu{NWwxjkf-feDHo-1}\nwixd{NWwxjkf-4V7g9k-1}}}%
\nwixlogsorted{c}{{check exactness and add to order}{NWwxjkf-8ug4W-1}{\nwixu{NWwxjkf-Dyu9X-1}\nwixd{NWwxjkf-8ug4W-1}}}%
\nwixlogsorted{c}{{check if node is exactly zero}{NWwxjkf-4Carso-1}{\nwixu{NWwxjkf-3j4HBX-1}\nwixd{NWwxjkf-4Carso-1}}}%
\nwixlogsorted{c}{{compute and add operation error}{NWwxjkf-1o2REi-1}{\nwixu{NWwxjkf-1F8iKo-1}\nwixd{NWwxjkf-1o2REi-1}}}%
\nwixlogsorted{c}{{constructors}{NWwxjkf-56Gik-1}{\nwixu{NWwxjkf-qfj8R-1}\nwixd{NWwxjkf-56Gik-1}}}%
\nwixlogsorted{c}{{destructor}{NWwxjkf-3BLOo7-1}{\nwixu{NWwxjkf-qfj8R-1}\nwixd{NWwxjkf-3BLOo7-1}}}%
\nwixlogsorted{c}{{factor in child unique degree}{NWwxjkf-3BswN0-1}{\nwixu{NWwxjkf-4cPL07-3}\nwixd{NWwxjkf-3BswN0-1}}}%
\nwixlogsorted{c}{{fixed precision computation}{NWwxjkf-1xhPU9-1}{\nwixu{NWwxjkf-qfj8R-1}\nwixd{NWwxjkf-1xhPU9-1}}}%
\nwixlogsorted{c}{{fixed precision routines}{NWwxjkf-1F8iKo-1}{\nwixu{NWwxjkf-1xhPU9-1}\nwixd{NWwxjkf-1F8iKo-1}}}%
\nwixlogsorted{c}{{fixed precision wrapper}{NWwxjkf-3EYhwl-1}{\nwixu{NWwxjkf-1xhPU9-1}\nwixd{NWwxjkf-3EYhwl-1}}}%
\nwixlogsorted{c}{{guaranteeing error}{NWwxjkf-xiB2Q-1}{\nwixu{NWwxjkf-qfj8R-1}\nwixd{NWwxjkf-xiB2Q-1}}}%
\nwixlogsorted{c}{{guaranteeing relative and absolute error}{NWwxjkf-Zc3P0-1}{\nwixu{NWwxjkf-xiB2Q-1}\nwixd{NWwxjkf-Zc3P0-1}\nwixd{NWwxjkf-Zc3P0-2}}}%
\nwixlogsorted{c}{{includes}{NWwxjkf-EOZR2-1}{\nwixu{NWwxjkf-qfj8R-1}\nwixd{NWwxjkf-EOZR2-1}}}%
\nwixlogsorted{c}{{increase operands reference counter}{NWwxjkf-4NmrBg-1}{\nwixu{NWwxjkf-56Gik-1}\nwixd{NWwxjkf-4NmrBg-1}}}%
\nwixlogsorted{c}{{increase single operand reference counter}{NWwxjkf-3ioqow-1}{\nwixu{NWwxjkf-56Gik-1}\nwixd{NWwxjkf-3ioqow-1}}}%
\nwixlogsorted{c}{{initialize bigfloat data}{NWwxjkf-4GuWSD-1}{\nwixu{NWwxjkf-qfj8R-1}\nwixd{NWwxjkf-4GuWSD-1}}}%
\nwixlogsorted{c}{{members}{NWwxjkf-3GwZ9F-1}{\nwixu{NWwxjkf-qfj8R-1}\nwixd{NWwxjkf-3GwZ9F-1}}}%
\nwixlogsorted{c}{{merge child vardeg nodes}{NWwxjkf-1hEUs5-1}{\nwixu{NWwxjkf-4cPL07-3}\nwixd{NWwxjkf-1hEUs5-1}}}%
\nwixlogsorted{c}{{mtdag\_node.hpp}{NWwxjkf-qfj8R-1}{\nwixd{NWwxjkf-qfj8R-1}}}%
\nwixlogsorted{c}{{node data constructors}{NWwxjkf-1sQxCF-1}{\nwixu{NWwxjkf-feDHo-1}\nwixd{NWwxjkf-1sQxCF-1}}}%
\nwixlogsorted{c}{{node data declaration}{NWwxjkf-feDHo-1}{\nwixu{NWwxjkf-qfj8R-1}\nwixd{NWwxjkf-feDHo-1}}}%
\nwixlogsorted{c}{{node data getter and setter}{NWwxjkf-1qdLgq-1}{\nwixu{NWwxjkf-qfj8R-1}\nwixd{NWwxjkf-1qdLgq-1}}}%
\nwixlogsorted{c}{{node data members}{NWwxjkf-Ai2gJ-1}{\nwixu{NWwxjkf-feDHo-1}\nwixd{NWwxjkf-Ai2gJ-1}}}%
\nwixlogsorted{c}{{node type}{NWwxjkf-2FFzRb-1}{\nwixu{NWwxjkf-qfj8R-1}\nwixd{NWwxjkf-2FFzRb-1}}}%
\nwixlogsorted{c}{{notify parents}{NWwxjkf-vBDTo-1}{\nwixu{NWwxjkf-2vFwl8-1}\nwixd{NWwxjkf-vBDTo-1}}}%
\nwixlogsorted{c}{{parent handling}{NWwxjkf-XH50e-1}{\nwixd{NWwxjkf-XH50e-1}\nwixd{NWwxjkf-XH50e-2}\nwixd{NWwxjkf-XH50e-3}\nwixd{NWwxjkf-XH50e-4}\nwixu{NWwxjkf-qfj8R-1}}}%
\nwixlogsorted{c}{{prepare synchronization}{NWwxjkf-2y2CB9-1}{\nwixu{NWwxjkf-3jZbw2-1}\nwixd{NWwxjkf-2y2CB9-1}}}%
\nwixlogsorted{c}{{propagate error topologically descending}{NWwxjkf-6vwGL-1}{\nwixu{NWwxjkf-1RZNmO-1}\nwixd{NWwxjkf-6vwGL-1}}}%
\nwixlogsorted{c}{{recompute approximation thread based}{NWwxjkf-3jZbw2-1}{\nwixu{NWwxjkf-1RZNmO-1}\nwixd{NWwxjkf-3jZbw2-1}}}%
\nwixlogsorted{c}{{reference handling}{NWwxjkf-2obXTf-1}{\nwixd{NWwxjkf-2obXTf-1}\nwixd{NWwxjkf-2obXTf-2}\nwixu{NWwxjkf-qfj8R-1}}}%
\nwixlogsorted{c}{{reset parameters}{NWwxjkf-26Om8R-1}{\nwixu{NWwxjkf-3j4HBX-1}\nwixd{NWwxjkf-26Om8R-1}\nwixu{NWwxjkf-4Carso-1}}}%
\nwixlogsorted{c}{{reset temporary parameters}{NWwxjkf-srI8w-1}{\nwixu{NWwxjkf-Dyu9X-1}\nwixd{NWwxjkf-srI8w-1}}}%
\nwixlogsorted{c}{{send notifications}{NWwxjkf-2vFwl8-1}{\nwixu{NWwxjkf-1RZNmO-2}\nwixd{NWwxjkf-2vFwl8-1}}}%
\nwixlogsorted{c}{{separation bound computation}{NWwxjkf-3rifsU-1}{\nwixd{NWwxjkf-3rifsU-1}\nwixu{NWwxjkf-qfj8R-1}}}%
\nwixlogsorted{c}{{separation bound computation compute degree}{NWwxjkf-4cPL07-1}{\nwixd{NWwxjkf-4cPL07-1}\nwixd{NWwxjkf-4cPL07-2}\nwixd{NWwxjkf-4cPL07-3}\nwixd{NWwxjkf-4cPL07-4}\nwixu{NWwxjkf-3rifsU-1}}}%
\nwixlogsorted{c}{{separation bound computation wrapper}{NWwxjkf-3PC6VY-1}{\nwixd{NWwxjkf-3PC6VY-1}\nwixu{NWwxjkf-3rifsU-1}}}%
\nwixlogsorted{c}{{separation bound concrete computation}{NWwxjkf-4Litu7-1}{\nwixu{NWwxjkf-3rifsU-1}\nwixd{NWwxjkf-4Litu7-1}}}%
\nwixlogsorted{c}{{separation bound getter and setter}{NWwxjkf-2Ingyi-1}{\nwixd{NWwxjkf-2Ingyi-1}\nwixu{NWwxjkf-1qdLgq-1}\nwixd{NWwxjkf-2Ingyi-2}}}%
\nwixlogsorted{c}{{set new error for binary operators}{NWwxjkf-2zgPbv-1}{\nwixu{NWwxjkf-ZVr1v-1}\nwixu{NWwxjkf-ZVr1v-2}\nwixd{NWwxjkf-2zgPbv-1}}}%
\nwixlogsorted{c}{{set new error for unary operators}{NWwxjkf-BA3wG-1}{\nwixu{NWwxjkf-ZVr1v-1}\nwixd{NWwxjkf-BA3wG-1}}}%
\nwixlogsorted{c}{{set p for addition and subtraction}{NWwxjkf-fTSXM-1}{\nwixu{NWwxjkf-ZVr1v-1}\nwixd{NWwxjkf-fTSXM-1}}}%
\nwixlogsorted{c}{{set q and check exactness}{NWwxjkf-1alRwG-1}{\nwixd{NWwxjkf-1alRwG-1}\nwixu{NWwxjkf-4EGrT7-1}\nwixu{NWwxjkf-ZVr1v-1}\nwixu{NWwxjkf-ZVr1v-2}}}%
\nwixlogsorted{c}{{sign computation}{NWwxjkf-VFFA-1}{\nwixu{NWwxjkf-qfj8R-1}\nwixd{NWwxjkf-VFFA-1}}}%
\nwixlogsorted{c}{{sign computation auxillary methods}{NWwxjkf-1M0F4U-1}{\nwixu{NWwxjkf-VFFA-1}\nwixd{NWwxjkf-1M0F4U-1}}}%
\nwixlogsorted{c}{{sign computation wrapper}{NWwxjkf-9mRa2-1}{\nwixu{NWwxjkf-VFFA-1}\nwixd{NWwxjkf-9mRa2-1}}}%
\nwixlogsorted{c}{{start new task}{NWwxjkf-2itq2c-1}{\nwixu{NWwxjkf-XH50e-4}\nwixd{NWwxjkf-2itq2c-1}}}%
\nwixlogsorted{c}{{static member definition}{NWwxjkf-2yBJ4v-1}{\nwixd{NWwxjkf-2yBJ4v-1}\nwixu{NWwxjkf-qfj8R-1}}}%
\nwixlogsorted{c}{{synchronize threads}{NWwxjkf-10iWX4-1}{\nwixu{NWwxjkf-3jZbw2-1}\nwixd{NWwxjkf-10iWX4-1}}}%
\nwixlogsorted{c}{{synchronize with main thread}{NWwxjkf-2U4UWm-1}{\nwixu{NWwxjkf-2vFwl8-1}\nwixd{NWwxjkf-2U4UWm-1}}}%
\nwixlogsorted{c}{{Thread\_manager.cpp}{NWwxjkf-47smee-1}{\nwixd{NWwxjkf-47smee-1}}}%
\nwixlogsorted{c}{{Thread\_manager.hpp}{NWwxjkf-47sd8i-1}{\nwixd{NWwxjkf-47sd8i-1}}}%
\nwixlogsorted{c}{{threading members}{NWwxjkf-3eVWfJ-1}{\nwixd{NWwxjkf-3eVWfJ-1}\nwixd{NWwxjkf-3eVWfJ-2}\nwixu{NWwxjkf-qfj8R-1}}}%
\nwixlogsorted{c}{{tm add task}{NWwxjkf-42cCML-1}{\nwixu{NWwxjkf-47sd8i-1}\nwixd{NWwxjkf-42cCML-1}}}%
\nwixlogsorted{c}{{tm conditioned launch new thread}{NWwxjkf-1cLBPM-1}{\nwixu{NWwxjkf-42cCML-1}\nwixd{NWwxjkf-1cLBPM-1}}}%
\nwixlogsorted{c}{{tm constructors and destructors}{NWwxjkf-2SHrAR-1}{\nwixu{NWwxjkf-47sd8i-1}\nwixd{NWwxjkf-2SHrAR-1}\nwixd{NWwxjkf-2SHrAR-2}}}%
\nwixlogsorted{c}{{tm debug includes}{NWwxjkf-3yauSd-1}{\nwixu{NWwxjkf-47sd8i-1}\nwixd{NWwxjkf-3yauSd-1}}}%
\nwixlogsorted{c}{{tm destroy thread}{NWwxjkf-AjUMk-1}{\nwixu{NWwxjkf-lCGr5-1}\nwixd{NWwxjkf-AjUMk-1}}}%
\nwixlogsorted{c}{{tm execute task}{NWwxjkf-fJsF4-1}{\nwixu{NWwxjkf-lCGr5-1}\nwixd{NWwxjkf-fJsF4-1}}}%
\nwixlogsorted{c}{{tm guard}{NWwxjkf-2q8kM5-1}{\nwixu{NWwxjkf-47sd8i-1}\nwixd{NWwxjkf-2q8kM5-1}}}%
\nwixlogsorted{c}{{tm guard instantiation}{NWwxjkf-3L7KKX-1}{\nwixu{NWwxjkf-4gZXMR-1}\nwixd{NWwxjkf-3L7KKX-1}}}%
\nwixlogsorted{c}{{tm includes}{NWwxjkf-2sbsH1-1}{\nwixu{NWwxjkf-47sd8i-1}\nwixd{NWwxjkf-2sbsH1-1}\nwixd{NWwxjkf-2sbsH1-2}\nwixd{NWwxjkf-2sbsH1-3}}}%
\nwixlogsorted{c}{{tm instantiation}{NWwxjkf-4gZXMR-1}{\nwixu{NWwxjkf-47sd8i-1}\nwixd{NWwxjkf-4gZXMR-1}}}%
\nwixlogsorted{c}{{tm launch thread if not max}{NWwxjkf-3BVqFP-1}{\nwixu{NWwxjkf-1cLBPM-1}\nwixd{NWwxjkf-3BVqFP-1}}}%
\nwixlogsorted{c}{{tm members}{NWwxjkf-23cSgs-1}{\nwixu{NWwxjkf-47sd8i-1}\nwixd{NWwxjkf-23cSgs-1}\nwixd{NWwxjkf-23cSgs-2}\nwixd{NWwxjkf-23cSgs-3}\nwixd{NWwxjkf-23cSgs-4}}}%
\nwixlogsorted{c}{{tm thread and task handling}{NWwxjkf-lCGr5-1}{\nwixu{NWwxjkf-47sd8i-1}\nwixd{NWwxjkf-lCGr5-1}\nwixd{NWwxjkf-lCGr5-2}\nwixd{NWwxjkf-lCGr5-3}}}%
\nwixlogsorted{c}{{topological evaluation getter and setter}{NWwxjkf-bCnDP-1}{\nwixu{NWwxjkf-1qdLgq-1}\nwixd{NWwxjkf-bCnDP-1}}}%
\nwixlogsorted{c}{{topological sorting}{NWwxjkf-Dyu9X-1}{\nwixd{NWwxjkf-Dyu9X-1}\nwixu{NWwxjkf-qfj8R-1}}}%
\nwixlogsorted{c}{{topsort visit children}{NWwxjkf-4SoTCX-1}{\nwixu{NWwxjkf-Dyu9X-1}\nwixd{NWwxjkf-4SoTCX-1}}}%
\nwixlogsorted{c}{{try to improve filter policy}{NWwxjkf-KEXze-1}{\nwixu{NWwxjkf-3j4HBX-1}\nwixu{NWwxjkf-3EYhwl-1}\nwixd{NWwxjkf-KEXze-1}}}%
\nwixlogsorted{c}{{typedefs and enums}{NWwxjkf-1oEtvZ-1}{\nwixu{NWwxjkf-qfj8R-1}\nwixd{NWwxjkf-1oEtvZ-1}}}%
\nwbegindocs{208}%

\end{document}